\documentclass[twocolumn, twocolappendix]{aastex631}

\usepackage{xspace}
\usepackage{threeparttable}
\usepackage{tabularx}
\usepackage{amsmath}

\usepackage{outlines}
\usepackage{float}
\usepackage[bottom]{footmisc}
\usepackage{xcolor}

\newcommand{\coone}{CO $J$=1-0\xspace}

\newcommand{\cothree}{CO $J$=3-2\xspace}

\newcommand{\velu}{km s$^{-1}$\xspace}
\newcommand{\Tkin}{$T_{\mathrm{kin}}$\xspace}
\newcommand{\voldens}{$n$\xspace}
\newcommand{\voldensu}{$\mathrm{cm^{-3}}$\xspace}
\newcommand{\Nco}{$N_{\mathrm{CO}}$\xspace}

\newcommand{\taucoone}{$\tau_{\mathrm{CO 1-0}}$\xspace}
\newcommand{\phibf}{$\Phi_{\mathrm{bf}}$\xspace}
\newcommand{\alphavir}{$\alpha_{\mathrm{vir}}$\xspace}
\newcommand{\alphaco}{$\alpha_{\mathrm{CO}}$\xspace}

\newcommand{\solarmass}{M\ensuremath{_{\odot}}\xspace}
\newcommand{\Sigmol}{$\Sigma_{\mathrm{mol}}$\xspace}

\newcommand{\coldenunit}{M$_{\odot}$ pc$^{-2}$\xspace}

\newcommand{\sfru}{$\mathrm{M_{\odot} yr^{-1}}$\xspace}

\submitjournal{ApJ}

\shorttitle{RADEX modeling for GMCs in the Antennae}
\shortauthors{He et al.}

\begin{document}

\title{Unraveling the Mystery of the Low CO-to-H$_2$ Conversion Factor in Starburst Galaxies: RADEX Modeling of the Antennae}

\author[0000-0001-9020-1858]{Hao He}
\affiliation{Argelander-Institut f\"ur Astronomie, Universit\"at Bonn, Auf dem H\"ugel 71, 53121 Bonn, Germany}
\affiliation{Department of Physics \& Astronomy, McMaster University, 1280 Main St. W., Hamilton, ON., L8S 4L8, Canada}

\author[0000-0001-5817-0991]{Christine D. Wilson}
\affiliation{Department of Physics \& Astronomy, McMaster University, 1280 Main St. W., Hamilton, ON., L8S 4L8, Canada}

\author[0000-0003-0378-4667]{Jiayi Sun}\altaffiliation{NASA Hubble Fellow}
\affiliation{Department of Astrophysical Sciences, Princeton University, 4 Ivy Lane, Princeton, NJ 08544, USA}
\affiliation{Department of Physics \& Astronomy, McMaster University, 1280 Main St. W., Hamilton, ON., L8S 4L8, Canada}

\author[0000-0003-4209-1599]{Yu-Hsuan Teng}
\affiliation{Center for Astrophysics and Space Sciences, Department of Physics, University of California San Diego, \\
9500 Gilman Drive, La Jolla, CA 92093, USA
}
\author[0000-0002-5204-2259]{Erik Rosolowsky}
\affiliation{Department of Physics, University of Alberta, Edmonton, AB T6G 2E1, Canada}

\author[0000-0003-0618-8473]{Ashley R. Bemis}
\affiliation{Department of Physics \& Astronomy, University of Waterloo, Waterloo, ON N2L 3G1, Canada}
\affiliation{Waterloo Centre for Astrophysics, University of Waterloo, 200 University Ave W, Waterloo, ON N2L 3G1, Canada}



\begin{abstract}

CO emission has been widely used as a tracer of molecular gas mass. However, it is a long-standing issue to accurately constrain the CO-to-H$_2$ conversion factor ($\alpha_{\mathrm{CO}}$) that converts CO luminosity to molecular gas mass, especially in starburst galaxies. We present the first resolved $\alpha_{\mathrm{CO}}$ modeling results with multiple ALMA CO and $^{13}$CO transition observations at both giant molecular cloud (GMC) scale at 150 pc and kpc scale for one of the closest starburst mergers, the Antennae. By combining our CO modeling results and measurements of 350 GHz dust continuum, we find that most GMCs in the Antennae have $\alpha_{\mathrm{CO}}$ values $\sim$4 times smaller than the commonly adopted Milky Way value (4.3). We find $\alpha_{\mathrm{CO}}$ at GMC scales shows a strong dependence on CO intensity, $^{13}$CO/CO ratio and GMC velocity dispersion, which is consistent with various theoretical and simulation predictions. Specifically, we suggest that the $^{13}$CO/CO line ratio and the velocity dispersion can be used to infer $\alpha_{\mathrm{CO}}$ in starburst regions. By applying our modeled $\alpha_{\mathrm{CO}}$ in GMC analyses, we find that GMCs in the Antennae are less gravitationally bound than in normal spiral galaxies, which is more consistent with what is predicted by merger simulations. At kpc scale, we find that our modeled $\alpha_{\mathrm{CO}}$ values are smaller than the modeled $\alpha_{\mathrm{CO}}$ at GMC scale by 40\%, which can be due to inclusion of a diffuse gas component with lower $\alpha_{\mathrm{CO}}$ values. We find a similar correlation of $\alpha_{\mathrm{CO}}$ and CO intensity at kpc scales to that at GMC scales.  

\end{abstract}



\section{Introduction} \label{sec:intro}


The cold and dense molecular gas in the interstellar medium (ISM) is the direct fuel for current and future star formation. Measuring the amount and properties
of the molecular gas is crucial for understanding star formation, the ISM, and their relations with galaxy evolution. Although H$_2$ is the dominant component of molecular gas, it is not normally observable due to the high excitation temperature ($T_{\mathrm{ex}}$) of its lines. Instead, the $^{12}$C${^{16}}$O $J$=1-0 line (hereafter CO $J$=1-0) is the most commonly used tracer for measuring the molecular gas mass. The CO-to-H2 conversion factor, $\alpha_{\mathrm{CO}}$ is commonly defined for the $J$ = 1-0 line as the ratio of total molecular gas to ($M_{\text{mol}}$ in \solarmass) to the CO $J$=1-0 luminosity ($L_{\mathrm{CO(1-0)}}$ in K km s$^{-1}$ pc$^2$), or equivalently, the ratio of molecular gas surface density (\Sigmol in \coldenunit) to the CO $J$=1-0 intensity ($I_{\mathrm{CO(1-0)}}$ in K km s$^{-1}$):
\begin{equation}
\alpha_{\mathrm{CO}} = \frac{M_{\text{mol}}}{L_{\mathrm{CO(1-0)}}} = \frac{\Sigma_{\text{mol}}}{I_{\mathrm{CO(1-0)}}} \left[\frac{\mathrm{M_{\odot}}}{\mathrm{K\ km\ s^{-1}\ pc^2}}\right]
\end{equation}
Given that CO is straightforward to observe, a concrete prescription for $\alpha_{\mathrm{CO}}$ as a function of local ISM properties has been a longstanding goal.

$\alpha_{\mathrm{CO}}$ was first calibrated for individual giant molecular clouds (GMCs) in our Milky Way based on virial methods \citep[e.g.][]{solomon1987, scoville1987, scoville1989, maloney1990, young1991}, optically thin tracers such as dust continuum \citep{boulanger1996, dame2001, planckcollaboration2011}, CO isotopologue lines \citep{goldsmith2008} and gamma-ray observations \citep[e.g.][]{strong1996, grenier2005, abdo2010}. These studies have found a nearly constant $\alpha_{\mathrm{CO}}$ around 4.3 $\mathrm{M_{\odot}\ (K\ km\ s^{-1}\ pc^{2})^{-1}}$ \citep[][ and references therein]{bolatto_co--h_2013} with scatter of 0.3 dex. However, $\alpha_{\mathrm{CO}}$ in the Central Molecular Zone (CMZ) can be 3 -- 10 times lower than the average value \citep[][and references therein]{bolatto_co--h_2013}. Furthermore, extra-galactic observations have found systematic variations of  up to one or two orders of magnitude across different galactic environments \citep[e.g.][]{bolatto2008, donovanmeyer2012, rebolledo2012, sandstrom_co--h2_2013}. This issue is further complicated by the fact that different calibration methods can lead to vastly discrepant estimates \citep[e.g. SMC][]{bolatto2003, leroy2011}. Therefore, assuming a constant $\alpha_{\mathrm{CO}}$ can introduce systematic bias in calculating molecular gas mass, surface density and related quantities, such as molecular gas depletion time, the cloud free-fall time, the virial parameter and the turbulent pressure \citep{sun_molecular_2022, sun2023}. 

Theoretical models and simulations suggest that $\alpha_{\mathrm{CO}}$ can depend on both small-scale GMC properties, such as temperature, volume and surface density \citep[][and references therein]{gong2020,hu2022}, and kpc-scale environmental properties, such as metallicity, galactic disk surface density \citep[e.g.][]{wolfire2010, narayanan2012, kazandjian2015, renaud_diversity_2019, hu2022}. Recently, a lot of progress has been made in calibrating the metallicity dependence \citep[e.g.][]{schruba2012, amorín2016, accurso2017}, which has been applied to several studies of CO emission in nearby galaxies \citep[e.g.][]{sun_molecular_2020, sun__dynamical_2020, pessa2021, sun2023}. However, we still lack  a general  prescription that incorporates all the related physical quantities at different scales. 

In particular, $\alpha_{\mathrm{CO}}$ is poorly constrained in starburst systems, such as ultra/ luminous infrared galaxies (U/LIRGs). Early studies \citep[e.g.][]{downes1993, bryant1996, bryant1999, solomon1997, downes_rotating_1998} find that $\alpha_{\mathrm{CO}}$ in U/LIRGs should be $\sim$ 4 times lower than the Milky Way value to give reasonable molecular gas mass values closer to dynamical mass estimates. Studies on a large sample of U/LIRGs using multi-CO line large velocity gradient (LVG) radiative transfer modeling \citep[e.g.][]{solomon2005, downes_rotating_1998, papadopoulos_molecular_2012} find consistent average  values around 1.1 $\mathrm{M_{\odot}\ (K\ km\ s^{-1}\ pc^{2})^{-1}}$ \citep[][with helium contribution]{downes_rotating_1998}. Therefore, a discrete bimodal $\alpha_{\mathrm{CO}}$ prescription or a modified version accounting for the deviation from the star-forming main sequence \citep[e.g.][]{magnelli2012, sargent_regularity_2014} is generally applied in observed normal spiral and starburst galaxies. However, there is likely a large galaxy-to-galaxy $\alpha_{\mathrm{CO}}$ variation for different U/LIRGs \citep[]{papadopoulos_molecular_2012, sliwa_extreme_2017, carleton2017a}, which is not captured by those $\alpha_{\mathrm{CO}}$ prescriptions. This problem is further complicated by recent works using optically thin tracers \citep[e.g.][]{dunne_dust_2022}, which suggest a Milky-Way like $\alpha_{\mathrm{CO}}$ value for these U/LIRGs. 

Besides galaxy-to-galaxy variation, theoretical works \citep{narayanan2012, bolatto_co--h_2013} also suggest that $\alpha_{\mathrm{CO}}$ could vary within galaxies depending on the local environment. \citet{narayanan_co-h2conversion_2011} suggest that the low $\alpha_{\mathrm{CO}}$ is caused by the increase in GMC temperature \citep[partly through thermal coupling with dust heated by UV radiation;][]{magnelli2012, olsen2016} and/or velocity dispersion \citep[out of self-gravity;][]{papadopoulos_molecular_2012}, which makes CO emission over-luminous. Recent galaxy merger observations \citep[e.g.][]{papadopoulos_molecular_2012} and simulations \citep[e.g.][]{bournaud2015} seem to favor the increase in velocity dispersion to play the major role. However, \citet{renaud_three_2019} show in their simulation that $\alpha_{\mathrm{CO}}$ is not a sole function of velocity dispersion and is also dependent on different merging stages. 

In order to diagnose GMCs physical states and reasons for $\alpha_{\mathrm{CO}}$ variation, it is necessary to observe multiple CO and other molecular lines (specifically optically thin lines) at GMC resolution ($\sim$ 100 pc) to perform comprehensive LVG modeling. 
This approach has recently been implemented across several nearby galaxy centers \citep[e.g.][]{teng2022,teng2023}. 
However, due to the limited sensitivity and resolution of current instruments, most LVG studies on individual starburst mergers \citep[e.g.][]{papadopoulos_molecular_2012, sliwa2012, sliwa2013, sliwa_around_2014, he2020} can only probe a limited number of gas-rich regions at kpc resolution, making it hard to extract any $\alpha_{\mathrm{CO}}$ dependence on GMC properties and local environments. 

As one of the closest starburst mergers, NGC 4038/9 (the Antennae) is an ideal target for this study. At a distance of 22 Mpc \citep{schweizer2008}, ALMA can readily resolve molecular gas at GMC scales. The total SFR of the Antennae is between 11 \sfru \citep[combining UV and 24 \textmu m tracing star formation 1 -- 400 Myr ago,][]{bemis2019} and 20 \sfru \citep[based on extinction corrected H$\alpha$ tracing SFR $\sim$ 1 -- 10 Myr ago,][]{Chandar2017}. The higher SFR value traced by H$\alpha$ is comparable to those of LIRGs, which suggests a starburst event that was just triggered recently several tens of Myr ago. As a typical major merger between two gas-rich galaxies, the Antennae has been well-studied in both simulations and observations. Most simulations \citep[e.g.][]{karl2010, privon_dynamical_2013, renaud_diversity_2019} suggest that the Antennae has just passed its second pericentric passage $\sim$ 40 Myr ago. Its central region hosts the two progenitor nuclei, still separated by about 7 kpc \citep{zhang_antennae_2001}. As a starburst merger, it also hosts a large number ($\sim$ 10$^4$) of young massive star clusters exceeding 10$^4$ \solarmass, with maximal mass reaching 10$^6$ \solarmass \citep[][]{whitmore2014a, mok_mass_2020, he2022}. The extreme number of YMCs will likely provide enough stellar feedback \citep{keller_superbubble_2014} to ultimately disperse the molecular gas and significantly reduce the $\alpha_{\mathrm{CO}}$ values \citep{renaud_diversity_2019}. 

In this paper, we perform LVG modeling on high-resolution ($\sim$ 150 pc) CO and $^{13}$CO molecular lines from ALMA observations of the Antennae to constrain the physical properties of the molecular gas and $\alpha_{\mathrm{CO}}$ at both GMC and kpc scales. In Section 2, we describe the observations and how we processed the data. In Section 3, we describe the RADEX modeling method that we used to derive gas physical quantities (e.g. temperature, volume density and CO column density) and $\alpha_{\mathrm{CO}}$. In Section 4, we present our modeled gas physical properties (e.g. kinetic temperature, volume density and CO column density) and their connection with different line ratios. In Section 5, we present our modeled $\alpha_{\mathrm{CO}}$ at GMC scale and compare its dependence on various GMC observational and physical quantities with theoretical, simulation and observational predictions. In Section 6, we apply our modeled $\alpha_{\mathrm{CO}}$ in calculation of GMC surface density and virial equilibrium states. In Section 7, we present modeled $\alpha_{\mathrm{CO}}$ at kpc scales and its comparison with $\alpha_{\mathrm{CO}}$ at GMC scales. We also explore the $\alpha_{\mathrm{CO}}$  dependence on kpc-scale gas properties (e.g. gas surface density, velocity dispersion and metallicity). The conclusions are summarized in Section 8. 

\section{Observations and Data Processing} \label{sec:data}

\subsection{ALMA Spectral Lines}
\label{subsec: ALMA_co}

\begin{table*}[htb!]
    \centering
    \caption{ALMA CO data products}
    \begin{threeparttable}
        \begin{tabularx}{\textwidth}{cccccccc}
        \hline \hline
         Data & Project & ALMA  & Native  & LAS  & Velocity  &  RMS$_{\mathrm{native}}$  & RMS$_{\mathrm{150 pc}}$ \\
         Type & ID & Band &  Resolution &  & Resolution & & \\
         (1) & (2) & (3) & (4) & (5) & (6) & (7) & (8) \\
         \hline
         CO $J$=1-0 & 2018.1.00272.S &  3 & $0\farcs84$, 90 pc & $14\farcm5$, 93 kpc & 2.54 \velu & 0.09 K & 0.05 K \\
         CO $J$=2-1 & 2018.1.00272.S & 6 & $0\farcs51$, 54 pc & $6\farcm9$, 44 kpc & 2.54 \velu & 0.24 K &  0.11 K \\
         \cothree & 2021.1.00439.S & 7 & $0\farcs67$, 71 pc  & $4\farcm5$, 29 kpc & 3.4 \velu & 0.09 K & 0.04 K\\
         $^{13}$CO $J$=1-0 & 2021.1.00439.S & 3 & $1\farcs41$, 150 pc & $15\arcmin$, 96 kpc & 2.7 \velu & 0.04 K & 0.04 K \\
         $^{13}$CO $J$=2-1 & 2018.1.00272.S & 6 & $0\farcs71$, 76 pc & $6\farcm9$, 44 kpc & 5.3 \velu & 0.09 K & 0.05 K \\
         continuum & 2021.1.00439.S  & 7 & $0\farcs65$, 72 pc & $0\farcm34$, 2.2 kpc  & -- & 0.18 mJy/beam & 0.4 mJy/beam \\ 
         \hline
     \end{tabularx}
     \begin{tablenotes}
         \item Columns: (1) CO spectral lines. (2) ALMA project ID. (3) ALMA observed frequency band. (4) Native resolution for the smallest round beam. (5) Largest angular scale. (6) Velocity resolution. (7) Noise of the image cubes at the native resolution. (8) Noise of the image cubes after smoothing to the resolution of 150 pc. 
     \end{tablenotes}
    \end{threeparttable}
    \label{tab:line_table}
\end{table*}

We use multiple CO lines (CO $J$=1-0, 2-1, 3-2 and $^{13}$CO $J$=1-0, 2-1) from the Atacama Large Millimeter/Submillimeter Array (ALMA) to determine the physical properties of the gas in the Antennae at 150 pc scale. We obtained ALMA Band 3, 6 and 7 observations from cycle 5 project 2018.1.00272.S and cycle 8 project 2021.00439.S to capture multiple CO and $^{13}$CO lines. A summary of the datasets is in Table \ref{tab:line_table}. For all five lines, we have observations from the 12 m-array used in both a compact and extended configuration, 7 m-array (ACA) and total power (TP) array to recover emission from $\sim$ 100 pc scale up to $\sim$ 50 kpc scale. The continuum does not have TP observations. Besides the lines listed in Table \ref{tab:line_table}, we also have C$^{18}$O $J$=1-0 and 2-1 detected in the same spectral tuning as $^{13}$CO $J$=1-0 and 2-1, respectively. 


We calibrate the raw visibility data with the observatory-supplied calibration scripts and the appropriate version of the CASA pipeline. From the calibrated measurement sets, we extract and image a relevant subset of visibility data for each molecular line using a modified version of the PHANGS--ALMA imaging pipeline \citep{leroy2021}. Before imaging the lines, we performed continuum subtraction by subtracting the 1st-order fit modeling on line-free channels. We then combine the 12m and 7m measurement sets together and perform the imaging. The imaging steps generally follow the PHANGS imaging scheme \citep{leroy2021}. 
For the weighting of the visibility data, we adopt the \textit{Briggs} method with robustness parameter of 0.5. After the cleaning, we then feather the cleaned image product with TP data and apply the primary beam correction to get final image cubes for each line. We also smooth all the images to the smallest round beam. In the final step, we  convert all the image cubes to units of Kelvin (K). 

\begin{figure*}
\centering
\gridline{
    \fig{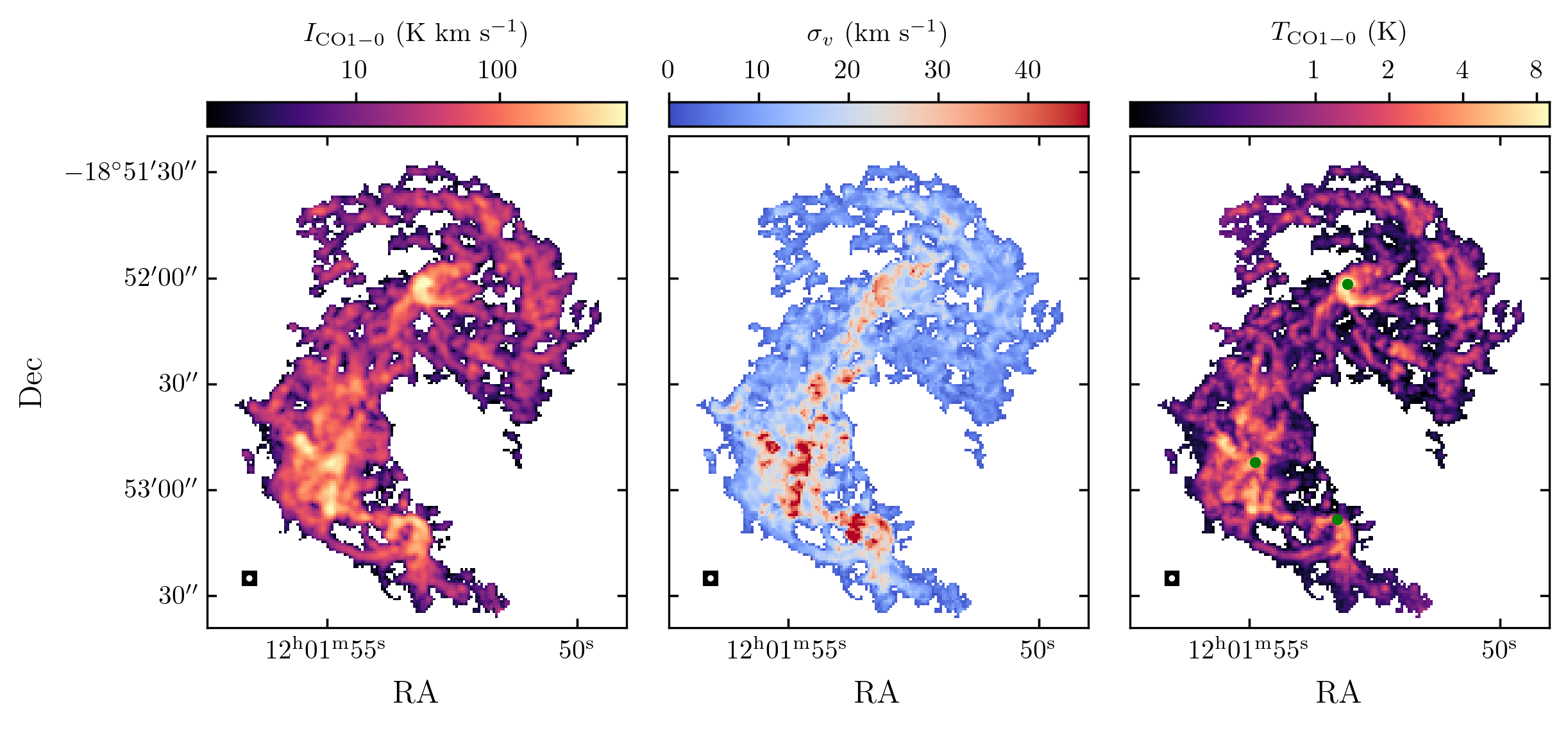}{0.9\textwidth}{}
	}
     \vspace{-2\baselineskip}
 \gridline{
    \fig{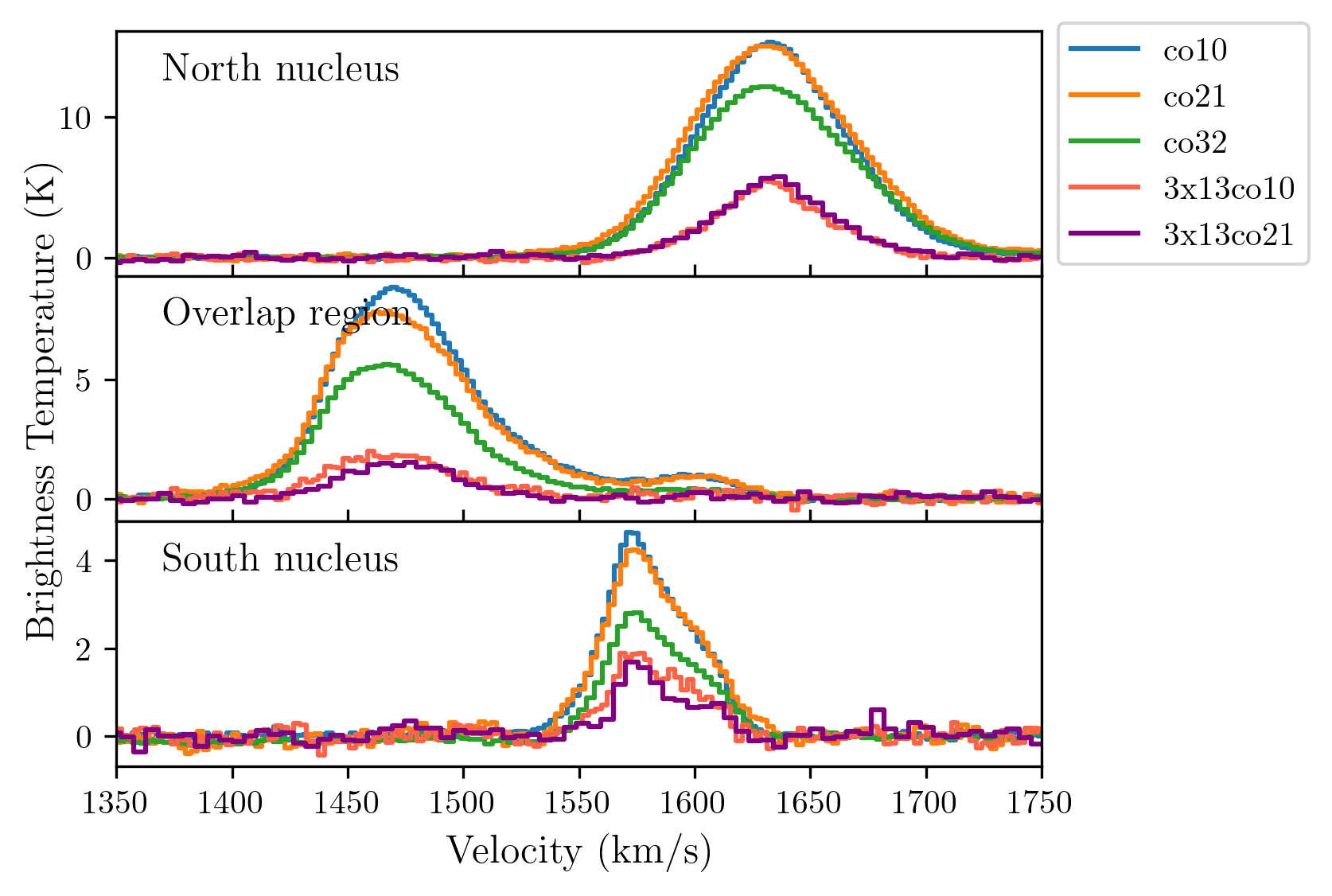}{0.6\textwidth}{}
 }
    \vspace{-2\baselineskip}
    \caption{(\textit{Top}) Integrated intensity, velocity dispersion and  peak brightness temperature of the CO $J$=1-0 observations at 150 pc resolution. The white circle within the black box is the beam of the image of 150 pc. The three green points in the map of peak brightness temperature map indicate three representative pixels in the north nucleus, overlap region and south nucleus. (\textit{Bottom}) CO and $^{13}$CO spectra for the three representative pixels marked in the top map.  }
    \label{fig:co10_moments}
\end{figure*}

\begin{figure*}
\centering
\vspace{-\baselineskip}
\gridline{
    \fig{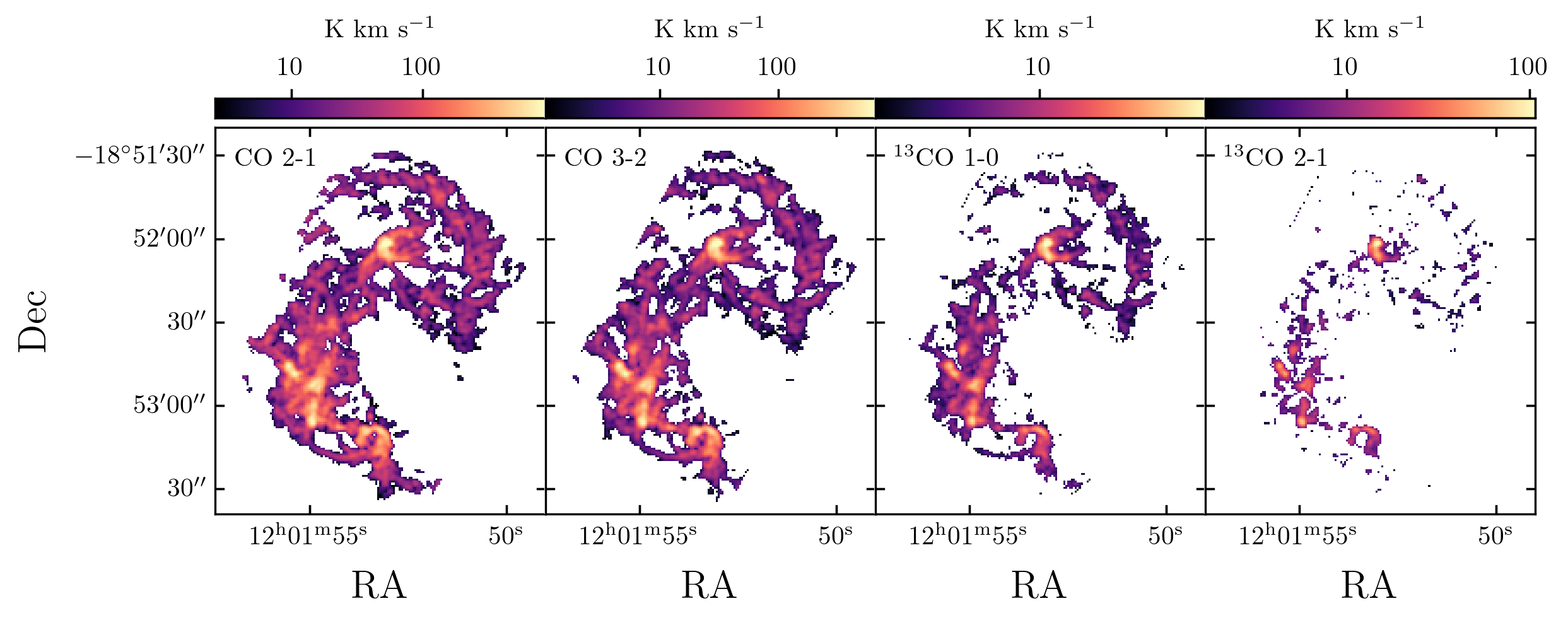}{\textwidth}{}
}
    \vspace{-2\baselineskip}
    \caption{Integrated intensity maps for CO $J$=2-1, 3-2 and $^{13}$CO $J$=1-0 and 2-1 lines at 150 pc resolution. Pixels with a S/N smaller than 3 are masked (see text in Section \ref{subsec: ALMA_co}).  }
    \label{fig:co_mom0}
\end{figure*}

We then perform post-processing steps to homogenize all 5 CO lines. We smooth all five image cubes to the resolution of 150 pc (1.41 arcsec) and match all their spatial grids to the CO $J$=1-0 line. We then produced a set of moment maps and effective width ($\sigma_v$) maps for all five lines at this common resolution. Specifically, the effective width is measured as the ratio between integrated intensity (moment 0) and peak brightness temperature (moment 8) maps, which is
\begin{equation}
\sigma_v = \frac{I}{\sqrt{2\pi}T_{\mathrm{peak}}}
\end{equation}
For a perfect Gaussian line profile, the effective width is identical to the traditionally used moment 2 measurements. We adopt this alternative method because it gives a more stable estimate of velocity dispersion within clouds, specifically if two or more clouds are along the same line of sight \citep[see][for more details]{heyer2001, sun_cloud-scale_2018}. To make moment maps, we start with generating masks adopting the scheme of the PHANGS-ALMA pipeline, which starts from a high-confidence mask including at least two consecutive channels with S/N above 5 and then expand the mask to include pixels with S/N above 2 for at least two consecutive channels. We run this scheme for each line and combine all the mask together to create a "combo" mask. We then apply this common "combo" mask to all the five line data cubes to make moment maps and their corresponding error maps. We also apply a S/N cut of 3 to the moment 0 maps of each line to exclude noisy pixels in weak line maps. In the final steps, we Nyquist-sample the moment 0 and effective width maps for all the lines to remove the spatial correlation between different pixels. Some representative moment maps are shown in Fig. \ref{fig:co10_moments} and \ref{fig:co_mom0}. 

In Section 7, we will also use the five CO and $^{13}$CO maps at a resolution of 1 kpc. We perform the same steps as described above to obtain the cubes and moment maps at this resolution. 

\subsection{ALMA continuum}

We also make the ALMA Band 7 continuum image in order to calculate the dust and gas mass. After the calibration of the Band 7 data, we use the PHANGS-ALMA pipeline \citep{leroy2021} to combine the 12m and 7m measurement sets and extract the line-free channels from the combined measurement set for the continuum imaging. We also collapse each spectral window into a single channel in order to speed up the continuum imaging process. We then use the \texttt{auto-multithresh} algorithm to clean the continuum data down to threshold of 2 $\times$ RMS (RMS of $\sim$ 0.18 mJy/beam). After imaging, we smooth the dust continuum map to the resolution of 150 pc and regrid the map to the nyquist-sampled CO images.  

\subsection{Spitzer Data}

We use the \textit{Spitzer} 3.6 \micron\ (Program 10136)  and 4.5 \micron\ (Program 61068) data to calculate the stellar mass surface density of the Antennae at kpc scale. We estimate the background level of each image by calculating the mean of an aperture drawn outside the galaxy and perform background subtraction. 
We reproject both images to match the pixel grids of the CO 150 pc resolution moment maps. We then calculate the stellar mass for each pixel with the equation \citep{eskew2012}
\begin{equation}
M_{\star} = 10^{5.65} F_{3.6}^{2.85} F_{4.5}^{-1.85} (D/0.05)^2 \, [\mathrm{M_{\odot}}], 
\end{equation}
where $F_{3.6}$ and $F_{4.5}$ are the flux at 3.6 and 4.5 \micron\ in Jy, respectively and $D$ is the luminosity distance in Mpc. We then calculate the stellar surface density by dividing by the pixel area using equation 
\begin{equation}
\Sigma_{\star} = M_{\star} / (150\ \text{pc})^2 \, [\mathrm{M_{\odot}\ pc^{-2}}]
\end{equation}
At the final step, we regrid our calculated stellar surface density map to the grid of 1 kpc resolution nyquist sampled CO $J$=1-0 moment 0 map. 

\begin{figure}
\centering
\vspace{-\baselineskip}
\gridline{
    \fig{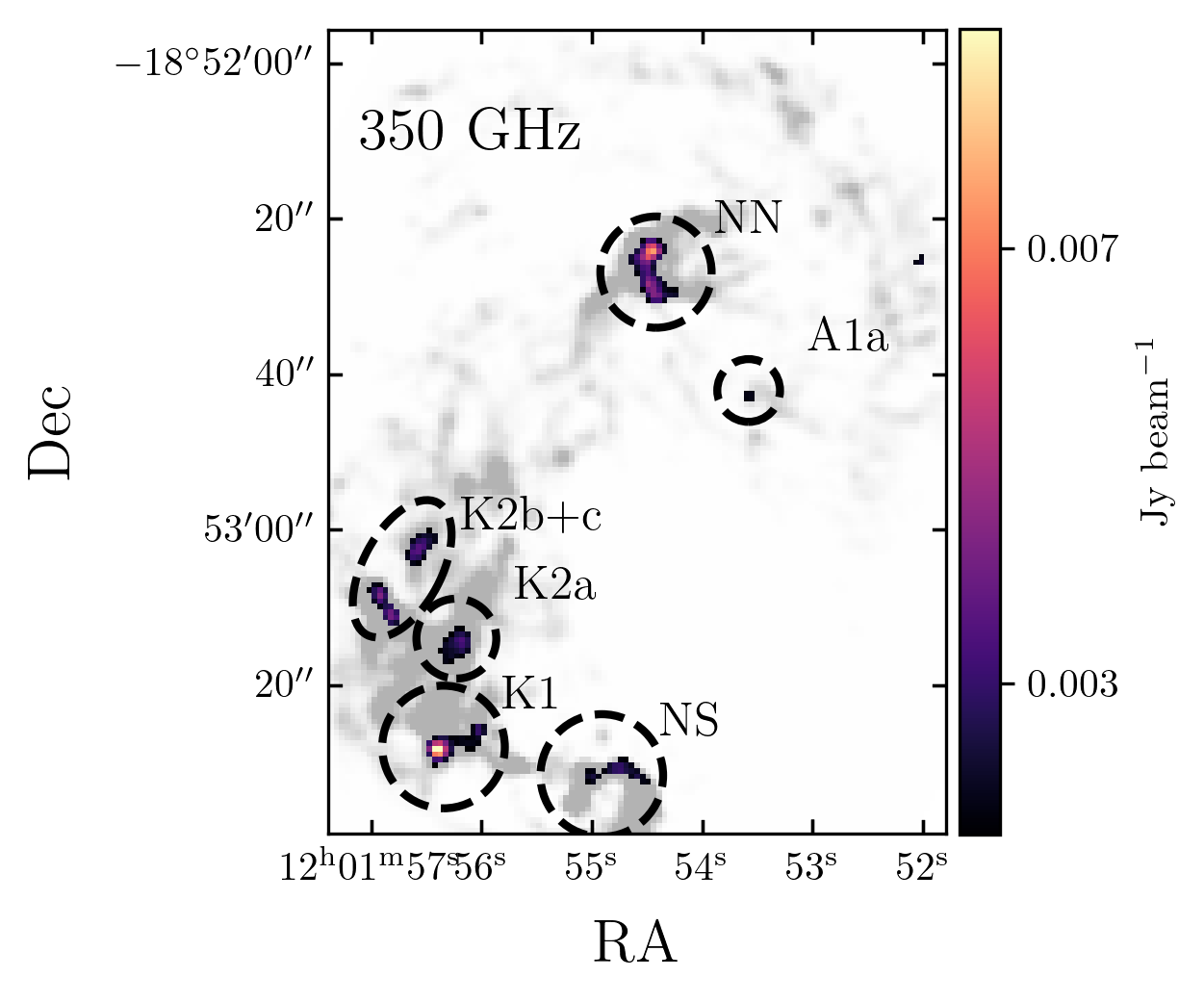}{0.4\textwidth}{}
}
    \vspace{-2\baselineskip}
    \caption{350 GHz dust continuum map overlaid on the CO $J$=1-0 integrated intensity map of the Antennae. Pixels with S/N $<$ 4 are masked. Dashed circles correspond to knots in the Herschel 70 \micron\ map identified in \citet{klaas2010}.  }
    \label{fig:dust_350GHz}
\end{figure}

\section{RADEX Modeling} \label{sec:RADEX_modeling}

\subsection{General Modeling Procedure}
\label{subsec:RADEX_general}

We adapt the code\footnote{https://github.com/ElthaTeng/multiline-bayesian-modeling \label{github}} from \citet{teng2022} to perform non-LTE radiative transfer modeling for each pixel with all five CO and $^{13}$CO lines detected at 150 pc. We briefly summarize the code and our adaptation below \citep[please refer to][for more details]{teng2022, teng2023}. This code runs RADEX \citep{van_der_tak_computer_2007} modeling, which assumes a homogeneous medium and uses radiative transfer equations based on the escape probability formalism to find a converged solution for the excitation temperature and level population for each molecular line transition. We adopt the one-component RADEX modeling to generate 5D grids of integrated line intensities for the five lines under combinations of varying H$_2$ volume density (\voldens), kinetic temperature (\Tkin), CO column density (\Nco), CO/$^{13}$CO abundance ratio ($X_{12/13}$) and beam filling factor (\phibf). Since RADEX modeling uses $N_{\mathrm{CO}}/\Delta v$ instead of \Nco alone, we instead sample CO column density per 15 \velu ($N_{\mathrm{CO}} \times \frac{15 \mathrm{km s^{-1}}}{\Delta v}$). The 15 \velu is the fiducial value used in \citet{teng2022, teng2023}. Later we will rescale this value to the real CO column density based on the measured $\Delta v$ in the CO $J$=1-0 velocity dispersion map ($\Delta v = 2.35 \sigma_v$). In the modeling, we assume the same \phibf for all 5 lines. We also assume the $[\mathrm{CO}]$/$[\mathrm{H}_2]$ abundance ratio ($x_{\mathrm{co}}$) of 3$\times 10^{-4}$. We will discuss our $x_{\mathrm{CO}}$ choice in Section \ref{subsec:dust_xco}. Our input parameters are summarized in Table 2. 

\begin{table}
\centering
\caption{RADEX Input Parameters}
\begin{threeparttable}
\begin{tabularx}{0.45\textwidth}{ccc}
    \hline \hline
     Parameter & Range & Step \\
     \hline
     log (\voldens) (\voldensu) & 2 -- 5.1 & 0.2   \\
     log (\Tkin) (K) & 1 -- 2.4 & 0.1 \\
     log ($N_{\mathrm{CO}} \times \frac{15 \mathrm{km s^{-1}}}{\Delta v}$) (cm$^{-2}$) & 16 -- 21 & 0.2 \\
     $X_{12/13}$ & 10 -- 400 & 10  \\
     \phibf & 0.05 -- 1 & 0.05 \\
     $[\mathrm{CO}]$/$[\mathrm{H}_2]$ & $3 \times 10^{-4}$ & -- \\
     \hline
\end{tabularx}
\end{threeparttable}
\end{table}

We then follow the Bayesian likelihood analyses in \citet{teng2022} to characterize the probability density function (PDF) for the five varying parameters (see detailed description in Appendix \ref{app:RADEX_stats}). From the derived 1D marginalized PDF, we can calculate the maximum likelihood ('1DMax') and median values for each variable. For all the parameters except for $X_{12/13}$, we use the median as the solution. For $X_{12/13}$, due to the bimodal shape of the 1D PDF, we instead use the 1DMax value as the solution (see detailed discussion in Appendix \ref{app:RADEX_stats}). 


\subsection{Modeling of the CO-to-H$_2$ conversion factor}
\label{subsec:RADEX_alphaCO}
The CO-to-H$_2$ conversion factor $\alpha_{\mathrm{CO}}$ is calculated \citep{teng2022} as 
\begin{equation}
\label{eq:alphaCO_def}
\begin{split}
\alpha_{\mathrm{CO}} &= \frac{\Sigma_{\text{mol}}}{I_{\mathrm{CO(1-0)}}} \left[\frac{\mathrm{M_{\odot}}}{\mathrm{K\ km\ s^{-1}}}\right] \\
&= \frac{1.36 m_{\mathrm{H_2}} N_{\mathrm{CO}} \Phi_{\mathrm{bf}}}{x_{\mathrm{co}} I_{\mathrm{CO(1-0)}}} \\
&= \frac{1}{4.5 \times 10^{19}} \frac{N_{\mathrm{CO}}[\mathrm{cm}^{-2}]\Phi_{\mathrm{bf}}}{x_{\mathrm{co}} I_{\mathrm{CO(1-0)}} [\mathrm{K\ km\ s^{-1}]}},    
\end{split}
\end{equation}
where $x_{\mathrm{co}}$ is the $[\mathrm{CO}]$/$[\mathrm{H}_2]$ abundance ratio. This equation includes the correction coefficient of 1.36 for the contribution of helium. This equation shows that the key modeling parameters to constrain $\alpha_{\mathrm{CO}}$ are the CO column density \Nco, beam filling factor \phibf and $[\mathrm{CO}]$/$[\mathrm{H}_2]$ abundance ratio.  

We then follow the method in \citet{teng2022} and \citet{teng2023} to derive the posterior probability of $\alpha_{\mathrm{CO}}$ (see detailed description in Appendix \ref{app:alphaCO}). For the rest of the paper, we choose the median value of the posterior as our derived $\alpha_{\mathrm{CO}}$, and 16$^{\mathrm{th}}$ and 84$^{\mathrm{th}}$ percentile as the $\pm 1\sigma$ values. We also inspect the marginalized distribution of the five modeled variables and $\alpha_{\mathrm{CO}}$ and exclude pixels with unreliable modeling results (see detailed description in Appendix \ref{app:alphaCO}). After our selection procedure, we have 508 pixels with good RADEX model constraints. 
 
Note that our derived $\alpha_{\mathrm{CO}}$ is dependent on our assumed $x_{\mathrm{CO}}$ value. The value we assume (3$\times 10^{-4}$) is commonly adopted for starburst systems \citep[e.g.][]{sliwa2017}. In Section \ref{subsec:dust_xco}, we further justify our $x_{\mathrm{CO}}$ choice by comparing the CO $J$=1-0 emission with the dust continuum. For comparison with studies with a different $x_{\mathrm{CO}}$ assumption, we give the equation to scale our $\alpha_{\mathrm{CO}}$ value based on different $x_{\mathrm{CO}}$ choices, which is 
\begin{equation}
\alpha_{\mathrm{CO}}^{\mathrm{scaled}} = \frac{3 \times 10^{-4}}{x_{\mathrm{CO}}} \alpha_{\mathrm{CO}}^{\mathrm{derived}}
\end{equation}

\section{Gas Physical Properties at GMC scales}

\subsection{Line Ratios}
\label{subsec:ratio}

\begin{figure*}
\centering
\gridline{
    \fig{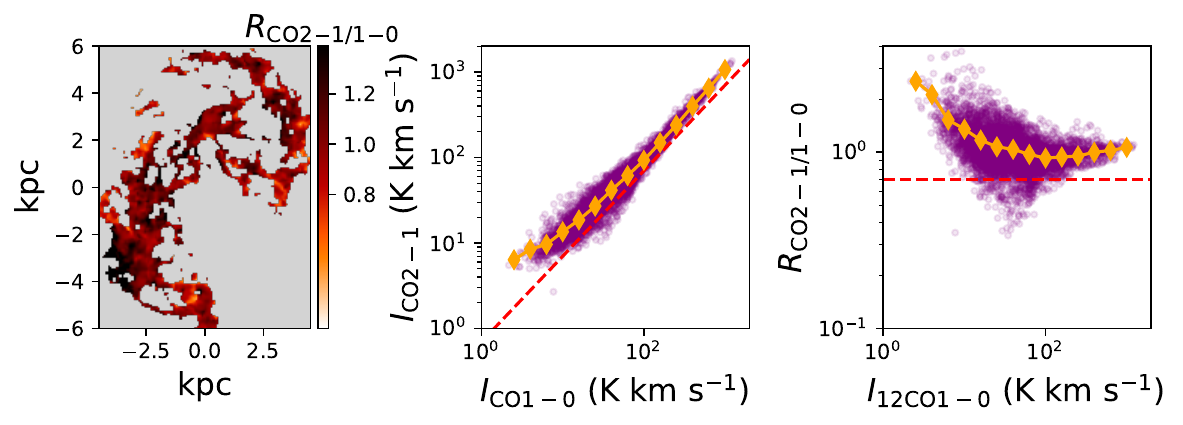}{\textwidth}{}
	}
 \vspace{-3\baselineskip}
\gridline{
    \fig{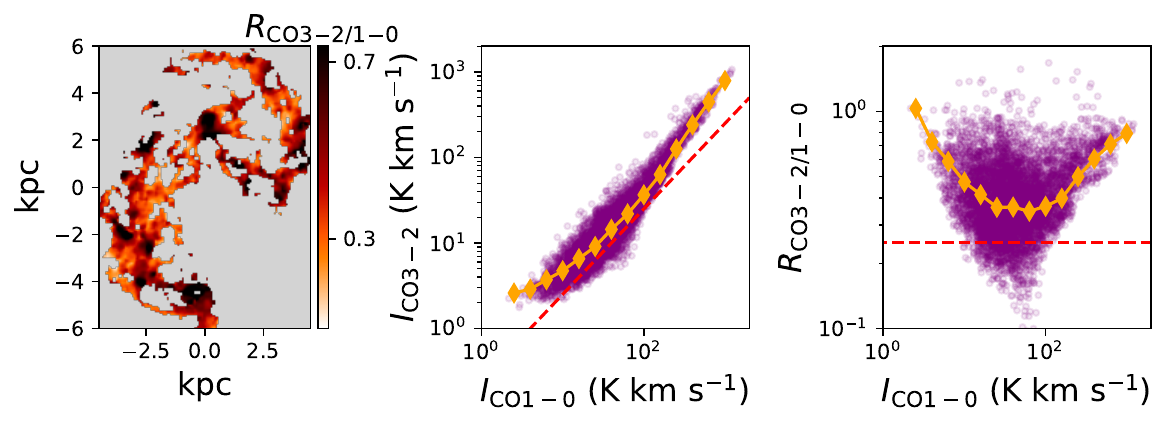}{\textwidth}{}
}
    \vspace{-2\baselineskip}
    \caption{Line ratios of ($\textit{upper}$) CO 2-1/1-0 and ($\textit{lower}$) CO 3-2/1-0. Orange diamonds specify the median for each CO 1-0 bin. The pixels are selected with S/N $>$ 5 for all three lines. The red dashed lines are the literature ratio values for normal spiral galaxies \citep[$R_{\mathrm{CO2-1/1-0}}=0.7$ and $R_{\mathrm{CO3-2/1-0}}=0.25$, ][]{sun_cloud-scale_2018}. We can see the line ratios in the Antennae are significantly higher than the literature values, which could be due to higher temperature or density of GMCs in starburst systems. The  $R_{\mathrm{CO2-1/1-0}}$ values in the Antennae are uniformly close to 1, which suggests both CO $J$=2-1 and 1-0 are thermally excited in regions that are warm and optically thick. On the other hand, $R_{\mathrm{CO3-2/1-0}}$ is significantly higher for regions with higher surface density, which could be either due to high gas temperature and/or volume density in these regions. The upward trend towards the low $I_{\mathrm{CO 1-0}}$ end for both ratios is due to the lower sensitivity of the CO$J$=2-1 and 3-2 lines (see footnote \ref{note:sensitivity})}
    \label{fig:co_ratio}
\end{figure*}

The \cothree/1-0 and CO $J$=2-1/1-0 ratios are indicators of the CO excitation, which is directly related to the molecular gas temperature and/or volume density \citep{leroy2017}. Fig. \ref{fig:co_ratio} shows the two ratio maps along with their dependencies on CO $J$=1-0 brightness temperature. We can see that the CO $J$=2-1/1-0 ratio is generally uniform with values close to 1 across the entire molecular gas detected regions. 
This ratio value is consistent with what is measured in starburst U/LIRGs \citep{montoyaarroyave2023}. This uniformity requires both lines to be thermally excited in regions that are warm and optically thick, which is expected to be typical of environments in starburst systems \citep[e.g.][]{sliwa2017}. Simulations also predict that the low-J CO lines are mostly thermalized for typical starburst mergers \citep[e.g.][]{bournaud2015}. Since most GMC observations for starburst mergers \citep[e.g.][]{brunetti_highly_2020, brunetti_extreme_2022} are done using the CO $J$=2-1 line due to its higher resolution and sensitivity, our study suggests that a typical starburst merger should have CO $J$=2-1/1-0 ratio values close to 1 instead of the commonly adopted 0.7 as seen in normal spiral galaxies \citep{leroy2021}. 

In contrast, the \cothree/1-0 ratio increases as the CO $J$=1-0 intensity (or gas surface density) increases (Fig. \ref{fig:co_ratio}, lower-right panel). This trend suggests that gas in these gas-concentrated regions is either denser and/or warmer than the rest of the regions. The average \cothree/1-0 ratio is $\sim$ 0.4 -- 0.7 (Fig. \ref{fig:co_ratio}, lower right panel), which is also consistent with typical U/LIRGs \citep[$\sim$ 0.76][]{montoyaarroyave2023} and the centers of normal spiral galaxies \citep{li2020, vlahakis2013}, but significantly higher than those in normal spiral galaxies \citep[$\sim$ 0.3][]{wilson2012, leroy2022}. This suggests that gas conditions of the Antennae are similar to those of gas concentrated and starburst environments in galaxy centers and U/LIRGs. 
\footnote{We also note that, for pixels with low CO intensities, both the CO $J$=2-1/1-0 and $J$=3-2/1-0 ratios increase towards the lower end. This is mainly due to the fact that the CO $J$=1-0 observation is much more sensitive than the CO $J$=2-1 and $J$=3-2 observations (Table \ref{tab:line_table}), and hence the two higher $J$ lines already hit the noise floor. \label{note:sensitivity}}

\begin{figure*}
\centering
\gridline{
    \fig{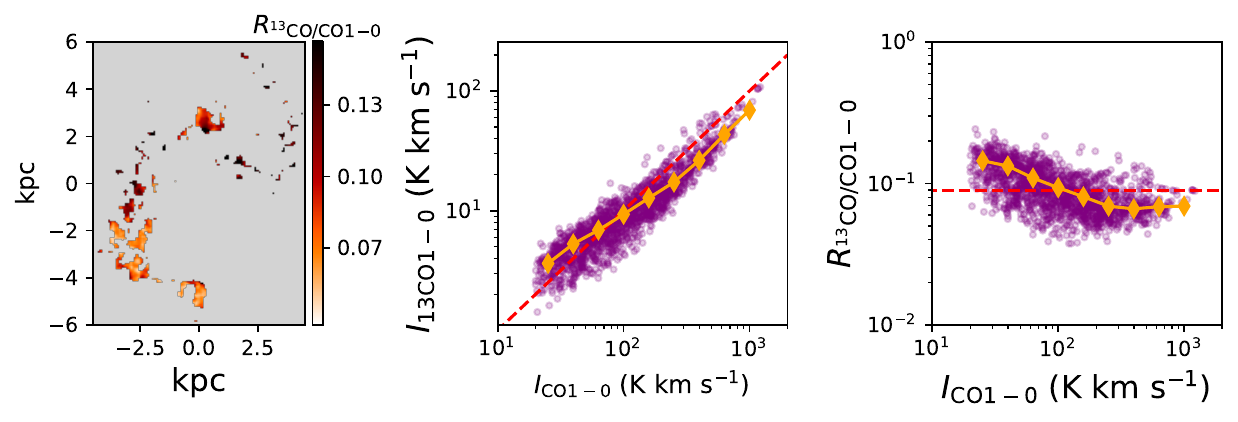}{\textwidth}{}
	}
 \vspace{-3\baselineskip}
\gridline{
    \fig{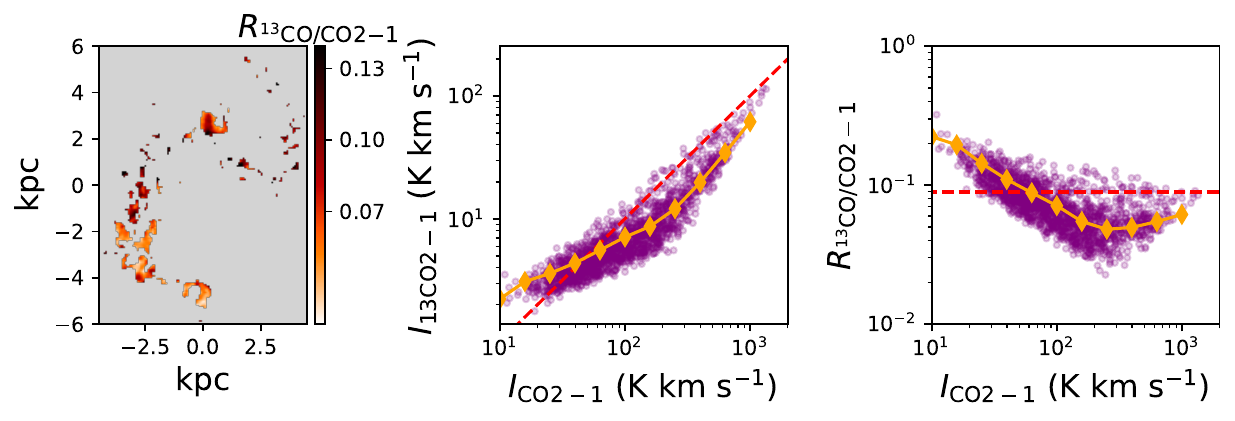}{\textwidth}{}
}
    \vspace{-2\baselineskip}
    \caption{Line ratios of $^{13}$CO/CO 1-0 ($\textit{upper}$) and 2-1 ($\textit{lower}$). The pixels are selected with S/N $>$ 3 for both $^{13}$CO lines. The red dashed lines are the literature ratio values \citep[$R_{\mathrm{^{13}CO/CO 1-0}}=0.09$][]{cormier_full-disc_2018}. The upward trend towards the low CO intensity end is probably due to the low sensitivity of the $^{13}$CO lines (see footnote \ref{note:sensitivity}). 
    }
    \label{fig:13co_ratio}
\end{figure*}

$^{13}$CO/CO $J$=1-0 and $^{13}$CO/CO $J$=2-1 ratio maps (Fig. \ref{fig:13co_ratio}) can be used to probe the [$^{13}$CO]/[CO] abundance ratio and the optical depth \citep[e.g.][]{jimenez-donaire_optical_2017}. Due to CO being optically thick, it is hard to disentangle these factors without comprehensive LVG modeling. To demonstrate this degeneracy, we consider a simple case where both CO and $^{13}$CO lines are thermally excited to the kinetic temperature. Under the local thermal equilibrium (LTE) condition, we would expect the $^{13}$CO/CO line ratio to be 
\begin{equation}
\begin{split}
R_{\mathrm{^{13}CO/CO 1{-}0}} &= \frac{T^{\mathrm{peak}}_{\mathrm{^{13}CO 1{-}0}}}{T^{\mathrm{peak}}_{\mathrm{CO 1{-}0}}}\\
&= \frac{\Phi_{\mathrm{bf}} T_{\mathrm{kin}} \left[1-\exp(-\tau_{\mathrm{^{13}CO 1{-}0}})\right]}{\Phi_{\mathrm{bf}} T_{\mathrm{kin}} \left[1-\exp(-\tau_{\mathrm{CO 1{-}0}})\right]} \\
& \approx \tau_{\mathrm{^{13}CO 1{-}0}}, \ (\tau_{\mathrm{^{13}CO 1{-}0}} \ll 1 \ll \tau_{\mathrm{CO 1-0}} ) \\
&= \tau_{\mathrm{CO 1{-}0}} / X_{12/13}
\end{split}
\label{eq:R10_tau13}
\end{equation}
where we assume the $^{13}$CO $J$=1-0 optical depth can be simply expressed as the CO $J$=1-0 optical depth divided by the ${[\mathrm{CO}]/[^{13}\mathrm{CO}]}$ abundance ratio ($X_{12/13}$). Therefore, a higher  $R_{\mathrm{^{13}CO/CO 1{-}0}}$ can be either due to higher CO optical depth (and hence higher column density) and/or lower $X_{12/13}$ abundance ratio. \footnote{We need to note that our simple derivation assumes both CO and $^{13}$CO $J$=1-0 lines are thermally excited and share the same beam filling factor $\Phi_{\mathrm{bf}}$ and the same linewidth $\Delta v$. In the real case, since CO $J$=1-0 is generally optically thick while $^{13}$CO $J$=1-0 is optically thin, the effective critical density of CO $J$=1-0 is lower than that of $^{13}$CO $J$=1-0 due to the line trapping effects. This will lead to lower beam filling factor and excitation temperature for the $^{13}$CO $J$=1-0 line and hence lower $R_{\mathrm{^{13}CO/CO 1{-}0}}$ value, specifically for lower-density regions \citep[see detailed discussion in][]{jimenez-donaire_optical_2017}. Therefore, we would also expect higher $R_{\mathrm{^{13}CO/CO 1{-}0}}$ in regions with higher gas volume density.} 

Both $R_{\mathrm{^{13}CO/CO 1{-}0}}$ and $R_{\mathrm{^{13}CO/CO 2{-}1}}$ have values of $\sim$ 0.1, similar to the typical $R_{\mathrm{^{13}CO/CO 1{-}0}}$ ratio for normal spiral galaxies \citep[e.g.][]{cormier_full-disc_2018}. On the other hand, this ratio is much higher than the typical ratio of starburst U/LIRGs \citep[$\sim$ 0.02][]{brown2019}. 
Since the Antennae has normal $^{13}$CO/CO line ratios but higher $X_{12/13}$ values ($\sim$ 200, Section \ref{subsec:results}) than normal spiral galaxies \citep[$\sim$ 60,][]{jimenez-donaire_optical_2017}, it is possible the normal $^{13}$CO/CO ratio is due to the combined effect of high optical depth and high $X_{12/13}$ values in the Antennae. We note that the Antennae has comparable molecular gas mass but less total SFR than typical U/LIRGs \citep[e.g. NGC 3256,][]{brunetti_cloud-scale_2022}, which might suggest that stellar feedback might not yet be effective to reduce the molecular gas optical depth in the Antennae. 


\subsection{Modeling results and their connection to the line ratios}
\label{subsec:results}

\begin{figure*}
\centering
\vspace{-\baselineskip}
\gridline{
    \fig{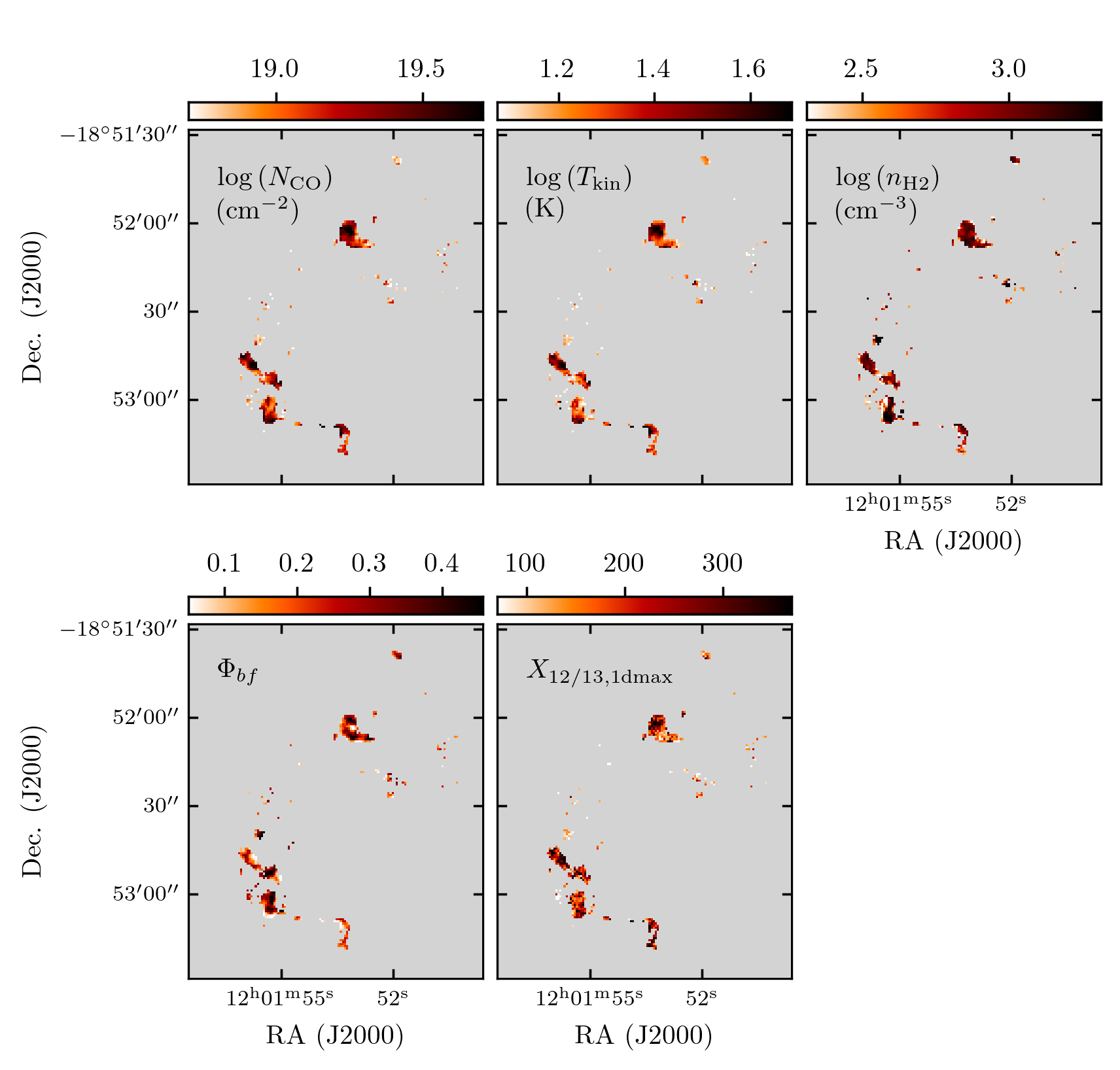}{0.9\textwidth}{}
}
    \vspace{-2\baselineskip}
    \caption{Maps of RADEX derived physical properties of the Antennae. The top row shows, from left to right, CO column density, kinetic temperature and molecular gas volume density. The bottom row shows the beam filling factor (assumed to be the same for all lines) on the left and the $[\mathrm{CO}]/[^{13}\mathrm{CO}]$ abundance ratio on the right. The maximal values for the color bars for all the quantities are set to be 95 percentile values. }
    \label{fig:results_maps}
\end{figure*}

\begin{figure*}
\centering
\vspace{-\baselineskip}
\gridline{
    \fig{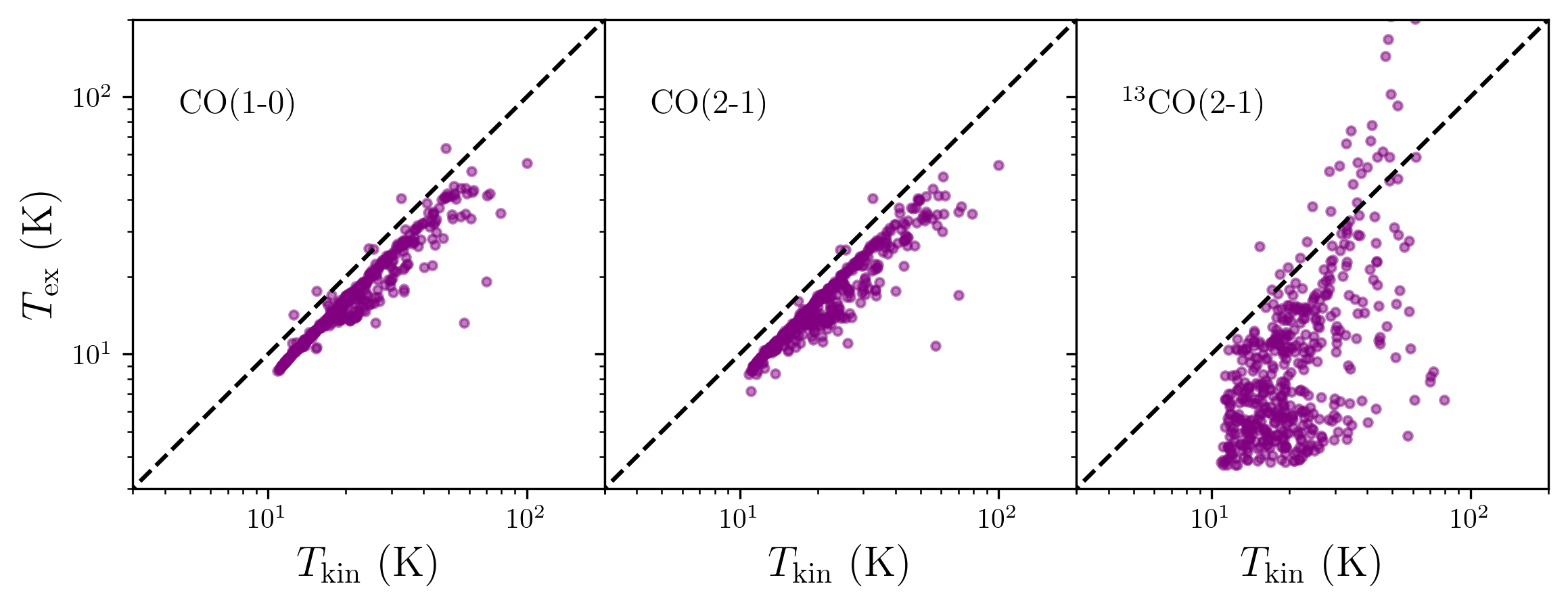}{0.9\textwidth}{}
}
\vspace{-2.5\baselineskip}
\gridline{
    \fig{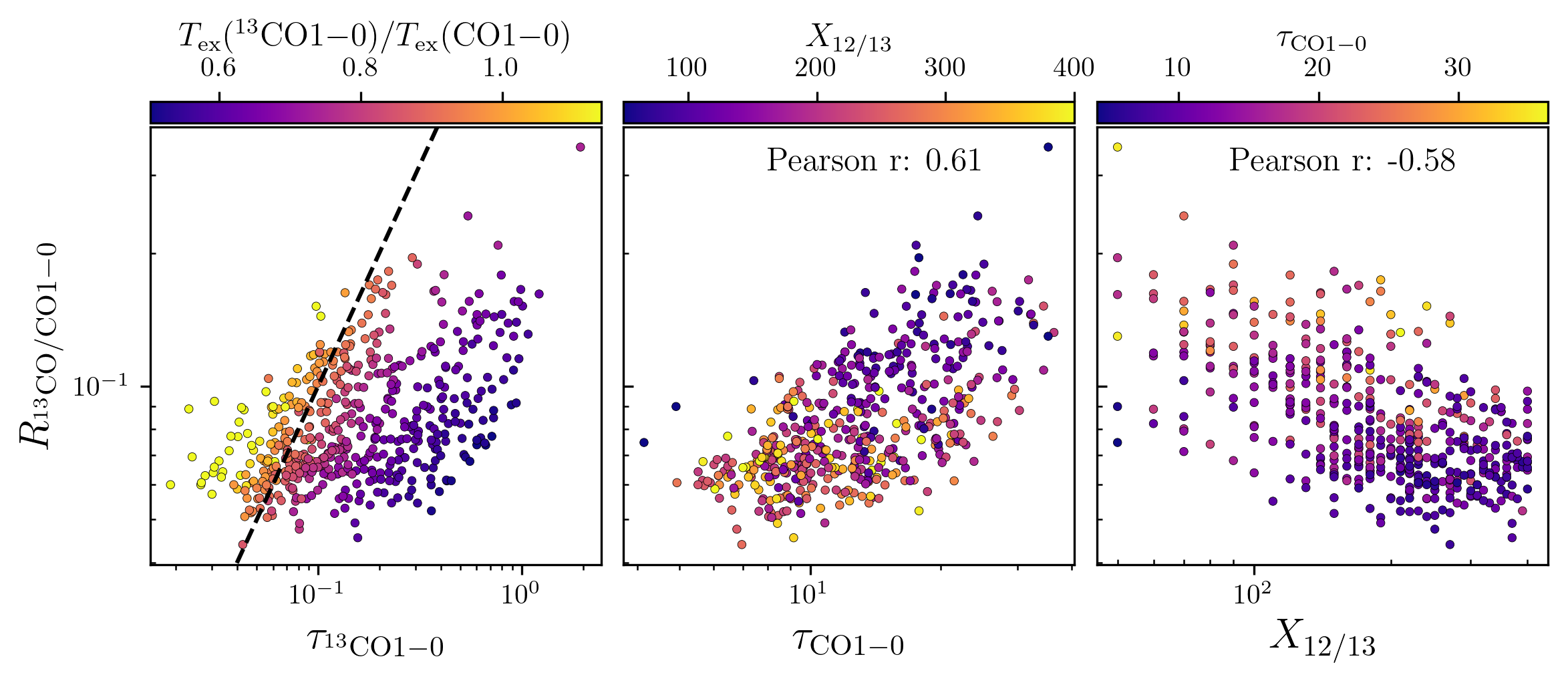}{0.9\textwidth}{}
}
    \vspace{-2\baselineskip}
    \caption{(\textit{Upper}) The excitation temperature of CO $J$=1-0, 2-1 and $^{13}$CO $J$=1-0 versus the kinetic temperature. We can see both CO $J$=1-0 and CO $J$=2-1 emissions are thermally excited (The excitation temperature is slightly lower than the kinetic temperature probably because the kinetic temperature is only sampled down to 10K, which affects the PDF of \Tkin and shift the median to slightly higher values). On the other hand, most $^{13}$CO $J$=1-0 emission is subthermally excited. (\textit{Lower left}) The $R_{\mathrm{^{13}CO/CO 1-0}}$ ratio versus $^{13}$CO $J$=1-0 optical depth. We can see when $^{13}$CO $J$=1-0 is thermally excited ($T_{\mathrm{ex}}(^{13}\mathrm{CO 1-0})/T_{\mathrm{ex}}(\mathrm{CO 1-0}) \approx 1$),  $R_{\mathrm{^{13}CO/CO 1-0}}$ ratio is roughly equal to the $^{13}$CO $J$=1-0 optical depth (dashed line), as expected under LTE conditions. (\textit{Lower middle}) $R_{\mathrm{^{13}CO/CO 1-0}}$ versus $\tau_{\mathrm{CO 1-0}}$ and (\textit{lower right}) $X_{12/13}$ abundance ratio. We can see $R_{\mathrm{^{13}CO/CO 1-0}}$ has a strong correlation with both quantities.}
    \label{fig:LTE_test}
\end{figure*}

We show the maps of our derived physical quantities in Fig. \ref{fig:results_maps}. We can see that most quantities show clear spatial variations among different regions. If we use the CO column density map as a guide, we can see that regions with higher gas surface densities, such as the two nuclei and the overlap region, generally have higher kinetic temperatures, beam filling factors and $[\mathrm{CO}]/[^{13}\mathrm{CO}]$ abundance ratios. On the other hand, the volume density distributions are more uniform throughout the entire galaxy\footnote{We note that we only include pixels with $^{13}$CO line detections for RADEX modeling. Therefore, we would expect these regions to have higher volume densities than those with only CO detections \citep{jimenez-donaire_optical_2017}.}.

The kinetic temperature map looks similar to the \cothree/1-0 ratio map. As previously discussed, the higher \cothree/1-0 ratio could either be caused by higher temperature or density. Since the volume density does not have as much spatial variation as the kinetic temperature, it seems the major driver for the \cothree/1-0 ratio variation is temperature. 

The CO $J$=2-1/1-0 ratio map looks uniformly close to 1, which suggests both lines are thermalized, with the excitation temperature equal to the kinetic temperature. Fig. \ref{fig:LTE_test} shows that the excitation temperatures of the CO $J$=2-1 and 1-0 lines are a good match to the kinetic temperature, which further confirms they are thermalized. Simulations \citep[e.g.][]{hu2022} suggest that for the CO $J$=1-0 and 2-1 lines, the LTE conditions are generally satisfied when kinetic temperature is above 10 K and volume density above 10$^3$ \voldensu. As we can see from the temperature and volume density maps (Fig. \ref{fig:results_maps}), most regions satisfy this condition. 

On the other hand, most $^{13}$CO $J$=1-0 emission is not thermally excited, with excitation temperature lower than the kinetic temperature. However, we expect that most regions should have volume density above the $^{13}$CO $J$=1-0 critical density \citep[650 \voldensu,][ Table 1]{jimenez-donaire_optical_2017} and will be thermally excited like CO $J$=1-0. However, our models assume $^{13}$CO $J$=1-0 has the same beam filling factor as the CO $J$=1-0 emission. In reality, $^{13}$CO $J$=1-0 might have smaller beam filling factor as it mostly comes from denser regions. In this case, we might overestimate the size and underestimate the actual excitation temperature of the $^{13}$CO $J$=1-0 emission. 

As discussed in Section \ref{subsec:ratio}, under LTE conditions, we would expect the $^{13}$CO/CO $J$=1-0 ratio to be equivalent to the $^{13}$CO $J$=1-0 optical depth. As shown in the lower-left panel of Fig. \ref{fig:LTE_test}, when LTE conditions hold for $^{13}$CO $J$=1-0 ($T_{\mathrm{ex}}(\mathrm{^{13}CO 1-0}) \approx T_{\mathrm{ex}}(\mathrm{CO 1-0})$), we see a 1-to-1 correspondence between the ratio and the optical depth. However, most regions have subthermally excited $^{13}$CO $J$=1-0 emission  according to our modeling, and the ratio in those regions is generally smaller than the actual $^{13}$CO $J$=1-0 optical depth. As discussed in Section \ref{subsec:ratio}, the two major drivers for $^{13}$CO/CO $J$=1-0 ratio variation are the CO $J$=1-0 optical depth and $X_{12/13}$ abundance ratio. Since $^{13}$CO $J$=1-0 might not satisfy LTE conditions, we test if the $^{13}$CO/CO $J$=1-0 ratio is still affected by those two factors using our RADEX modeling results. As shown in bottom middle and left panel of Fig. \ref{fig:LTE_test}, our RADEX modeling results suggest that $R_{\mathrm{^{13}CO/CO 1-0}}$ still has strong correlations with both quantities.

Fig. \ref{fig:results_maps} shows a clear spatial variation of $X_{12/13}$ that corresponds well with \Tkin and \Nco variation. Regions with higher CO column density and kinetic temperature generally have higher $X_{12/13}$. In theory, the high $X_{12/13}$ value could be caused by starburst activity that generates large amounts of $^{12}\mathrm{C}$ at short timescales \citep[$\sim$ 10 Myr][]{vigroux1976} by massive stars while $^{13}\mathrm{C}$ will only be released by intermediate-mass stars after $\sim$ 1 Gyr. This scenario is consistent with our expectation since regions with higher gas temperature and surface densities are generally where starburst events happen. In Section \ref{subsec:ratio}, we also discussed the negative correlation between $R_{\mathrm{^{13}CO/CO 1-0}}$ and CO $J$=1-0 intensity. The low $R_{\mathrm{^{13}CO/CO 1-0}}$ in high surface density regions could be potentially due to high $X_{12/13}$ in these regions. We also find most regions have $X_{12/13}$ values between 100 -- 300. This value is more similar to the $X_{12/13}$ in starburst mergers \citep[125 in Arp 220, 250 in NGC 2623][]{sliwa_pdbi_2017, sliwa2017} and significantly higher than that in the solar neighborhood and normal disk galaxies ($\sim$ 70) \citep[e.g.][]{langer1990, milam2005}. These high $X_{12/13}$ values are also consistent with the scenario that starburst events boost the $X_{12/13}$ ratio. 

Finally, we note that our modeling results are limited to regions with $^{13}$CO $J$=1-0 and $J$=2-1 detections. The maps in Fig. \ref{fig:results_maps} show that most detections are in the two nuclei and the overlap regions. To expand our modeling to other regions with lower gas surface densities, we need $^{13}$CO observations, specifically $^{13}$CO $J$=2-1 observations, with better sensitivity. Alternatively, we can stack $^{13}$CO pixels based on CO $J$=1-0 brightness temperature to achieve a better S/N level for the CO faint regions. 


\section{The CO-to-H$_2$ conversion factor at GMC scales}
\label{subsec:alphaCO_GMC_dependence}

\subsection{Dependence on CO $J$=1-0 intensity}
\label{subsubsec:alphaCO_Ico}


\begin{figure*}
\centering
\gridline{
    \fig{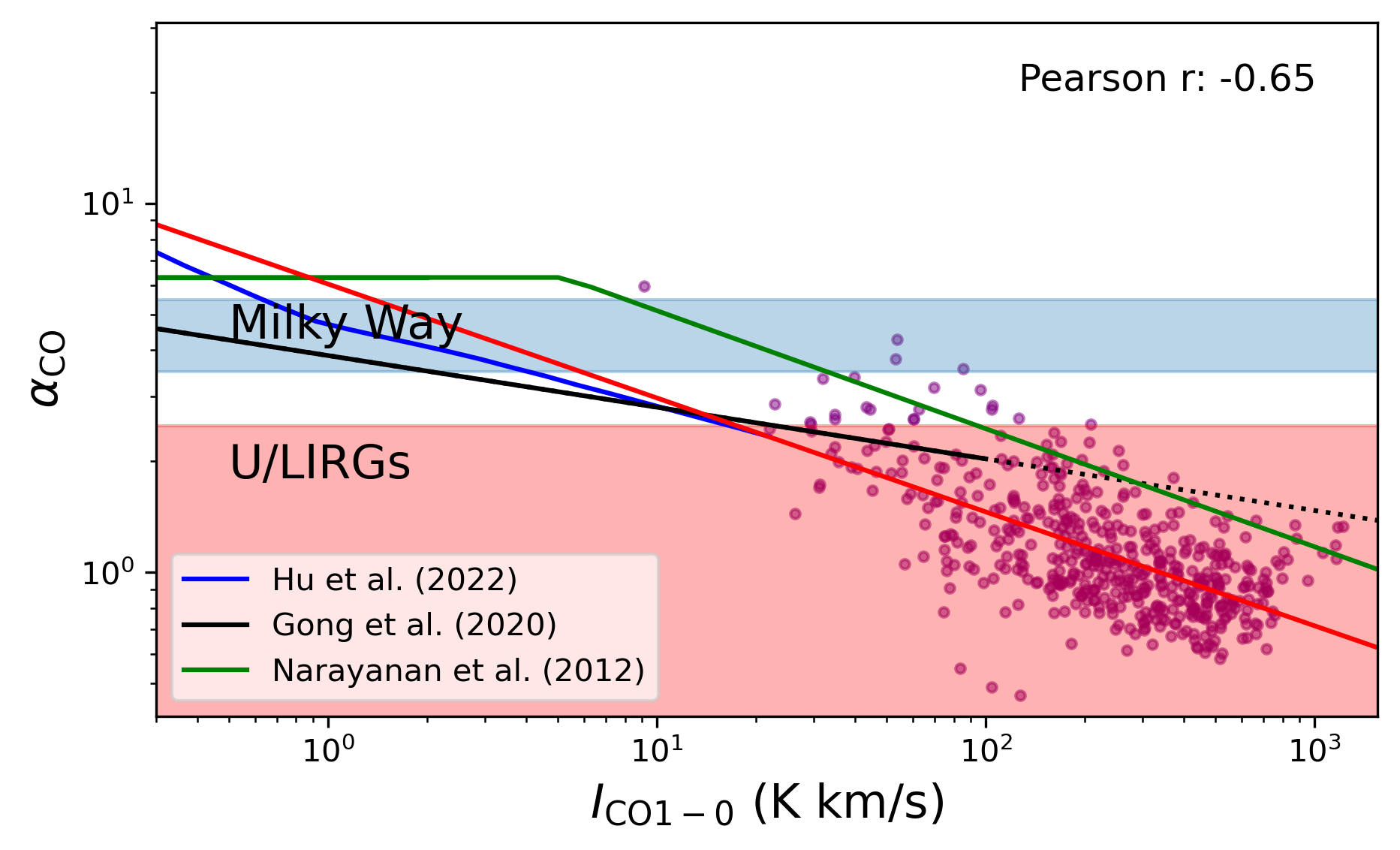}{0.7\textwidth}{}
	}
    \vspace{-2\baselineskip}
    \caption{(Modeled $\alpha_{\mathrm{CO}}$ versus CO $J$=1-0 integrated intensity $I_{\mathrm{CO(1-0)}}$. The red solid line is the fit to $\alpha_{\mathrm{CO}}$ versus $I_{\mathrm{CO(1-0)}}$ using data in this work (slope of -0.3). The black solid line is the simulation prediction from \citet{gong2020} at a resolution of 128 pc (slope of -0.14), the blue solid line is the simulation prediction from \citet{hu2022} at a resolution of 125 pc and the green solid line is the simulation prediction from \citet[][slope of -0.3]{narayanan2012}. The dotted lines are extrapolated relations out of the corresponding simulation ranges. 
    We can see a significant anti-correlation between $\alpha_{\mathrm{CO}}$ and $I_{\mathrm{CO(1-0)}}$ with a power-law slope closest to the prediction from \citet{narayanan2012}, which has an environment similar to starburst mergers such as the Antennae. The difference in power-law slope between other simulation predictions and the observational fit might be because we are probing an environment with much higher gas surface density.}
    \label{fig:alphaCO_Ico}
\end{figure*}

Various simulations \citep[e.g.][]{narayanan2012} have proposed that the bimodal distribution of $\alpha_{\mathrm{CO}}$ across normal spiral galaxies and U/LIRGs can be accounted for by an $\alpha_{\mathrm{CO}}$ that is anti-correlated with CO $J$=1-0 intensity ($I_{\text{CO 1-0}}$) at kpc scales. Recent simulations \citep[e.g.][]{gong2020, hu2022} have further pushed this dependence down to GMC spatial scales. Our study provides the first direct test of this dependence in starburst mergers. Fig. \ref{fig:alphaCO_Ico} shows a significant anti-correlation between $\alpha_{\mathrm{CO}}$ and $I_{\text{CO 1-0}}$. Our fit power-law relation is
\begin{equation}
\log \alpha_{\mathrm{CO}} = 0.77 (\pm 0.04) - 0.3 (\pm 0.02) \log I_{\mathrm{CO 1-0}}
\end{equation}
The value in the brackets indicates the 1-$\sigma$ error for the fitted parameter. We can see that our power-law fit has almost the same slope as that predicted (slope of -0.32) in \citet{narayanan2012}. On the other hand, the predictions from \citet{gong2020} and \citet{hu2022} give shallower slopes and tend to over-estimate $\alpha_{\mathrm{CO}}$ values. However, since the simulations of both \citet{gong2020} and \citet{hu2022} have a maximum $I_{\text{CO 1-0}}$ barely reaching 100 K km s$^{-1}$, their predictions are more applicable to the environment in normal spiral galaxies with much lower gas surface densities. 

The simulation run in \citet{narayanan2012} is focused on the molecular gas in starburst mergers with similar environments as the Antennae. Despite similar slopes, the predicted $\alpha_{\mathrm{CO}}$ values based on the relation in \citet{narayanan2012} are higher than our $\alpha_{\mathrm{CO}}$ measurement by a factor of 1.7. It is possible that this difference is caused by different $x_{\mathrm{CO}}$ values adopted in \citet{narayanan2012}. In their simulation, only part of the carbon is converted into CO molecules and the fraction depends on UV field strength, metallicity and optical extinction. In contrast, our models assume all the carbon is converted to CO, which gives a higher $x_{\mathrm{CO}}$ value, and hence lower $\alpha_{\mathrm{CO}}$ value. We further justify our $x_{\mathrm{CO}}$ choice in Section \ref{subsec:dust_xco}. To test which $x_{\mathrm{CO}}$ prescription is true, we need to test the relation in \citet{narayanan2012} in a larger sample of U/LIRGs at cloud-scale resolution.

\subsection{Dependence on CO 1-0 optical depth and $^{13}$CO/CO ratio}
\label{subsubsec:alphaCO_tau}

\begin{figure*}
\vspace{-1\baselineskip}
\centering
\gridline{
    \fig{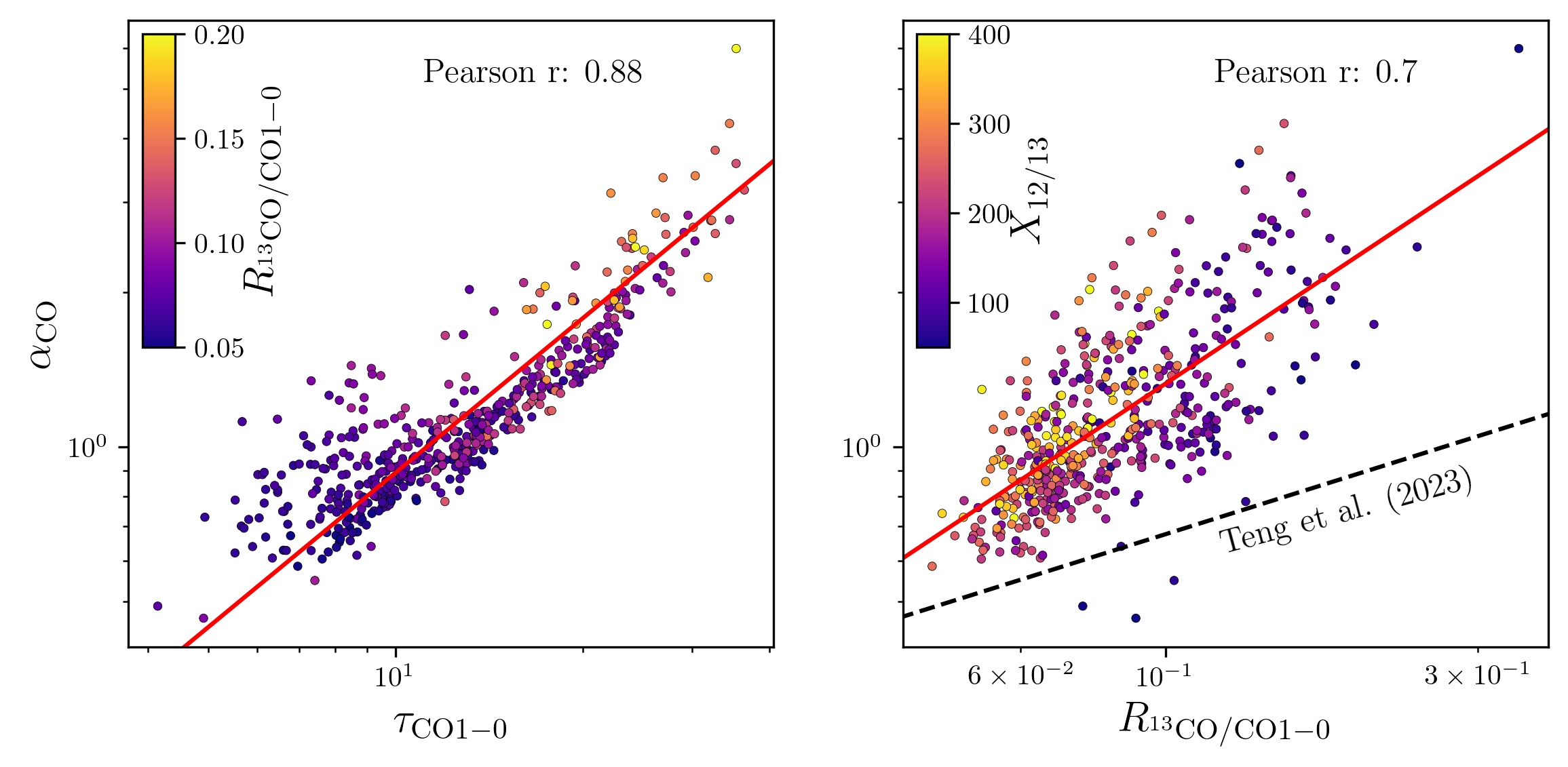}{0.9\textwidth}{}
	}
    \vspace{-2\baselineskip}
    \caption{(\textit{Left}) The modeled $\alpha_{\mathrm{CO}}$ versus the modeled $\tau_{\mathrm{CO(1-0)}}$ color coded by $^{13}$CO/CO $J$=1-0 ratio ($R_{^{13}\mathrm{CO/CO 1-0}}$). The red line is a proportional fit (power law slope of 1) to the data points. (\textit{Right}) Modeled $\alpha_{\mathrm{CO}}$ versus $R_{^{13}\mathrm{CO/CO 1-0}}$ color coded by [CO]/[$^{13}$CO] abundance ratio $X_{12/13}$. The red line is the power-law fit to the $\alpha_{\mathrm{CO}}$ versus  $R_{^{13}\mathrm{CO/CO 1-0}}$ relation (slope of 0.85). The dashed line is the $\alpha_{\mathrm{CO}}$ versus $R_{^{13}\mathrm{CO/CO 2-1}}$ relation from \citet[][slope of 0.4]{teng2023}. We can see there is a tight linear correlation between $\alpha_{\mathrm{CO}}$ and \taucoone for the Antennae, which is consistent with the results of \citet{teng2022,teng2023}. We can also see a strong correlation between $\alpha_{\mathrm{CO}}$ and $R_{^{13}\mathrm{CO/CO 1-0}}$, which suggests $R_{^{13}\mathrm{CO/CO 1-0}}$ can be potentially used as an $\alpha_{\mathrm{CO}}$ tracer.  }
    \label{fig:alphaCO_tau}
\end{figure*}

Recent GMC LVG modeling in normal spiral galaxies \citep{teng2022, teng2023} suggests that $\alpha_{\mathrm{CO}}$ has a tight proportional correlation with the CO $J$=1-0 optical depth (\taucoone) when $\tau_{\mathrm{CO(1-0)}} > 1$. This relation is consistent with our LVG modeling expectation, for which
\begin{equation}
\label{eq:optical_depth}
\tau_{\mathrm{CO 1-0}} \propto N_{\mathrm{CO}}/\Delta v   
\end{equation}
As shown in Section \ref{subsec:ratio}, most CO $J$=1-0 emission is thermally excited, giving
\begin{equation}
T_{\mathrm{peak}}  \approx \Phi_{\mathrm{bf}} T_{\mathrm{kin}} \left[1-\exp(-\tau_{\mathrm{CO 1-0}})\right]
\end{equation}
Substituting these two equations into Eq. \ref{eq:alphaCO_def}, we obtain
\begin{equation}
\label{eq:alphaCO_tau}
\begin{split}
\alpha_{\mathrm{CO}} &= \frac{N_{\mathrm{CO}} \Phi_{\mathrm{bf}}}{x_{\mathrm{co}} I_{\mathrm{CO(1-0)}}} = \frac{N_{\mathrm{CO}} / \Delta v}{x_{\mathrm{co}} T_{\mathrm{peak}}} \Phi_{\mathrm{bf}}  \\
 & \propto \frac{\tau_{\mathrm{CO 1-0}}}{1-\exp(-\tau_{\mathrm{CO 1-0}})},  \quad (x_{\mathrm{CO}}, T_{\mathrm{kin}} = \mathrm{const})  \\
 & \propto \tau_{\mathrm{CO 1-0}},  \quad (\tau_{\mathrm{CO 1-0}} \gg 1) 
\end{split}
\end{equation}
The left panel of Fig. \ref{fig:alphaCO_tau} shows a tight linear correlation between $\alpha_{\mathrm{CO}}$ and \taucoone, consistent with this theoretical expectation. The red solid line is the proportional fit to the relation, which is 
\begin{equation}
\label{eq:alphaCO_tau}
\log \alpha_{\mathrm{CO}} = \log \tau_{\mathrm{CO 1-0}} - 1.05 (\pm 0.01) 
\end{equation}

In real observations, $\tau_{\mathrm{CO 1-0}}$ is hard to measure and generally requires multi-CO transition radiative transfer modeling. Various studies \citep[e.g.][]{jimenez-donaire_optical_2017} have proposed to use the $^{13}$CO/CO line ratio ($R_{\mathrm{^{13}CO/CO}}$) to trace the optical depth of the optically thick CO $J$=1-0 line. Based on Eq. \ref{eq:R10_tau13}, we would expect $R_{\mathrm{^{13}CO/CO 1{-}0}} \propto \tau_{\mathrm{CO 1-0}}$ if the $X_{12/13}$ abundance ratio is fixed. This scenario is also supported by \citet{teng2023}, who found a strong correlation between $\alpha_{\mathrm{CO}}$ and $R_{\mathrm{^{13}CO/CO}}$ based on J=2-1 lines across three barred galaxy centers. As shown in the right panel of Fig. \ref{fig:alphaCO_tau}, we can see a strong correlation between $\alpha_{\mathrm{CO}}$ and $R_{\mathrm{^{13}CO/CO 1{-}0}}$. This strong correlation suggests $R_{\mathrm{^{13}CO/CO 1{-}0}}$ is a good indicator of \taucoone and hence can be used to infer $\alpha_{\mathrm{CO}}$. We fit a power-law function between these two quantities and have
\begin{equation}
 \log \alpha_{\mathrm{CO}} = 0.85 (\pm 0.04) \log R_{\mathrm{^{13}CO/CO 1{-}0}} + 0.97 (\pm 0.04)     
\end{equation}
We can see that the $\alpha_{\mathrm{CO}}$ versus $R_{\mathrm{^{13}CO/CO 1{-}0}}$ relation is still relatively close to linear as we expect. However, this relation is steeper than the $\alpha_{\mathrm{CO}}$ versus $R_{\mathrm{^{13}CO/CO 2{-}1}}$ relation obtained in \citet{teng2023}. This difference could potentially be due to the choice of different excitation lines (as discussed in Appendix \ref{sec:ratio_sensitivity}). We also note the difference in absolute values between our prescription and that in \citet{teng2023}, which could be due to their overestimate of $x_\mathrm{CO}$ (as suggested by \citealt{teng2024}) or the different galactic environments we study.  

The $\alpha_{\mathrm{CO}}$ versus $R_{\mathrm{^{13}CO/CO 1{-}0}}$ relation has a significantly larger scatter compared to the $\alpha_{\mathrm{CO}}$ versus \taucoone relation. We find that the scatter mainly comes from the varying $X_{12/13}$ value. As shown in the right panel of Fig. \ref{fig:alphaCO_tau}, higher $\alpha_{\mathrm{CO}}$ values correspond to higher $X_{12/13}$ values for a fixed $R_{\mathrm{^{13}CO/CO 1{-}0}}$ value. To account for this variation, we also perform a two-variable fitting between $\alpha_{\mathrm{CO}}$ and $R_{\mathrm{^{13}CO/CO 1{-}0}}$ and $X_{12/13}$, which is 
\begin{equation}
\label{eq:alphaCO_R13to12}
\begin{split}
\log \alpha_{\mathrm{CO}} &=  1.03 (\pm 0.05) \log R_{\mathrm{^{13}CO/CO 1-0}} \\
&+ 0.21 (\pm 0.03) \log X_{12/13} + 0.7 (\pm 0.06)
\end{split}
\end{equation}
We note that $X_{12/13}$ is not a direct observable. For Milky Way studies, $X_{12/13}$ can be measured through line ratios of optically thin lines \citep[e.g. $\mathrm{C^{18}O/{13}C^{18}O}$][]{langer1990}, which are hard to detect in extragalactic observations. Instead, the $X_{12/13}$ value for starburst U/LIRGs is constrained through RADEX modeling \citep[e.g.][Arp 220, NGC 6240, Arp 55 and NGC 2623]{sliwa_pdbi_2017, sliwa2017}. These systems show a large variation in the global $X_{12/13}$ value. For these systems with known global $X_{12/13}$ values, Eq. \ref{eq:alphaCO_R13to12} might be a better choice to infer $\alpha_{\mathrm{CO}}$ and hence GMC surface density. 


\subsection{Dependence on GMC velocity dispersion}
\label{subsubsec:alphaCO_alphavir}

As an optically thick line, CO $J$=1-0 is often established as a molecular gas tracer based on the fact that the CO $J$=1-0 luminosity is proportional to virial mass for individual GMCs in both the Milky Way and nearby galaxies \citep[references in][]{bolatto_co--h_2013}. However, GMCs in starburst systems might have larger velocity dispersion as perturbed by starburst/merging activity and hence be less gravitationally bound. For a given fixed surface density, the increase in velocity dispersion could reduce the optical depth of the gas (Eq. \ref{eq:optical_depth}) and hence reduce the $\alpha_{\mathrm{CO}}$ (Eq. \ref{eq:alphaCO_tau}). Early theoretical works \citep[e.g.][]{downes1993} suggested that for starburst systems, CO instead traces the geometric mean of molecular gas mass and virial mass \citep[i.e. $L_{\mathrm{CO}} \propto T_{\mathrm{B}, 0}\left(M_{\mathrm{gas}} M_{\mathrm{vir}} / \rho_{\mathrm{gas}}\right)^{1 / 2}$, see discussions in ][]{shetty2011}. \citet{papadopoulos_molecular_2012} infer the dynamical states of GMCs in starburst U/LIRGs based on the LVG modeled volume density and velocity gradient. Their results are also consistent with recent simulation predictions \citep[e.g.][]{bournaud2015}, who find that the major cause of low $\alpha_{\mathrm{CO}}$ in starburst mergers is due to  higher velocity dispersion instead of higher gas temperature in these systems compared to other types of galaxies. Recent high-resolution ALMA observations \citep{teng2023, teng2024} suggest a strong anti-correlation between $\alpha_{\mathrm{CO}}$ and GMC velocity dispersion also exists in the center of spiral galaxies.

\begin{figure*}
\centering
\gridline{
    \fig{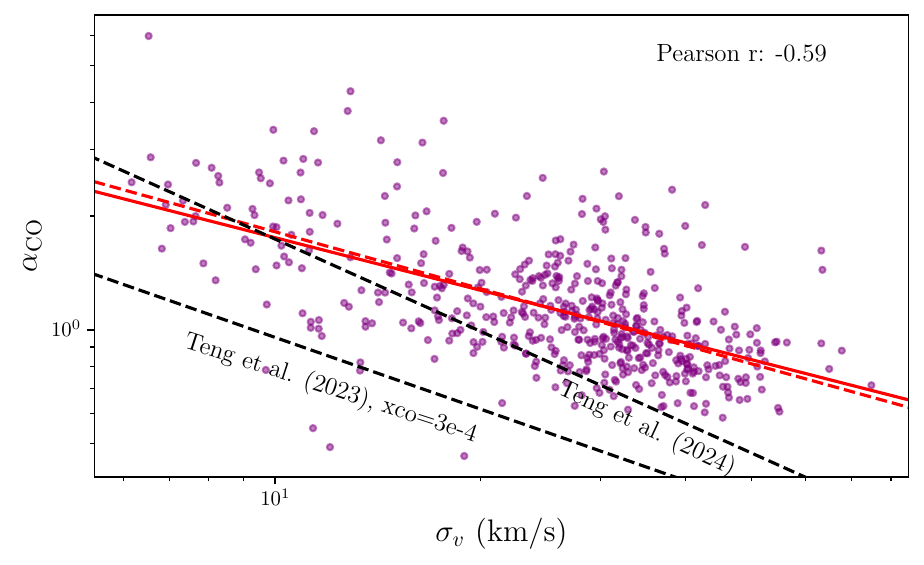}{0.7\textwidth}{}
	}
    \vspace{-2\baselineskip}
    \caption{$\alpha_{\mathrm{CO}}$ versus velocity dispersion $\sigma_v$. The red solid line is the power-law fit of the relation and the red dashed line is the power-law fit with fixed slope of -0.5 from theoretical predictions (see Section \ref{subsubsec:alphaCO_alphavir} for detailed discussion). The two black dashed lines indicate the relations found in \citet{teng2023, teng2024}. We can see a strong anti-correlation between $\alpha_{\mathrm{CO}}$ and $\sigma_v$ consistent with theoretical prediction, which suggests that the increase in velocity dispersion is responsible for bringing down $\alpha_{\mathrm{CO}}$ in starburst galaxy mergers. }
    \label{fig:alphaCO_vdep}
\end{figure*}

In Fig. \ref{fig:alphaCO_vdep}, we see a significant anti-correlation between $\alpha_{\mathrm{CO}}$ and $\sigma_v$, consistent with our modeling expectation. We fit the anti-correlation with a power-law function, which is
\begin{equation}
\log \alpha_{\mathrm{CO}} = 0.7 (\pm 0.04) - 0.46 (\pm 0.03) \log \sigma_{v}
\end{equation}
as well as with a power law function with a fixed slope of -0.5. This power-law slope is derived from theoretical predictions under the LVG approximation for two-level optically thick systems (see detailed discussion in \citet{teng2024} and references therein). 
Our fit power-law function is quite close to the theoretical prediction, which suggests that the variation in velocity dispersion could be the major driver to the $\alpha_{\mathrm{CO}}$ variation. Furthermore, this result suggests that we can use velocity dispersion to calibrate the $\alpha_{\mathrm{CO}}$ variation at GMC scales.

Fig. \ref{fig:alphaCO_vdep} also shows the $\alpha_{\mathrm{CO}}$ versus $\sigma_v$ fit from \citet{teng2023, teng2024}. We can see our data align relatively well with the calibrated relation in \citet{teng2024} using a dust-based approach \citep{sandstrom2013}, but are significantly higher than the calibrated relation in \citet{teng2023}. \citet{teng2024} suggest that the discrepancy between their calibrated $\alpha_{\mathrm{CO}}$ versus $\sigma_v$ relations from LVG modeling and the dust-based approach might be due to their adoption of a higher $x_{\mathrm{CO}}$ in the LVG modeling. As discussed in Section \ref{subsec:dust_xco}, our adopted value of $x_{\mathrm{CO}} = 3 \times 10^{-4}$ is reasonable for the Antennae, but might be too large for the centers of the normal spiral galaxies studied by \citet{teng2023}.

\subsection{Dependence on CO line ratios}

\begin{figure*}
\centering
\gridline{
    \fig{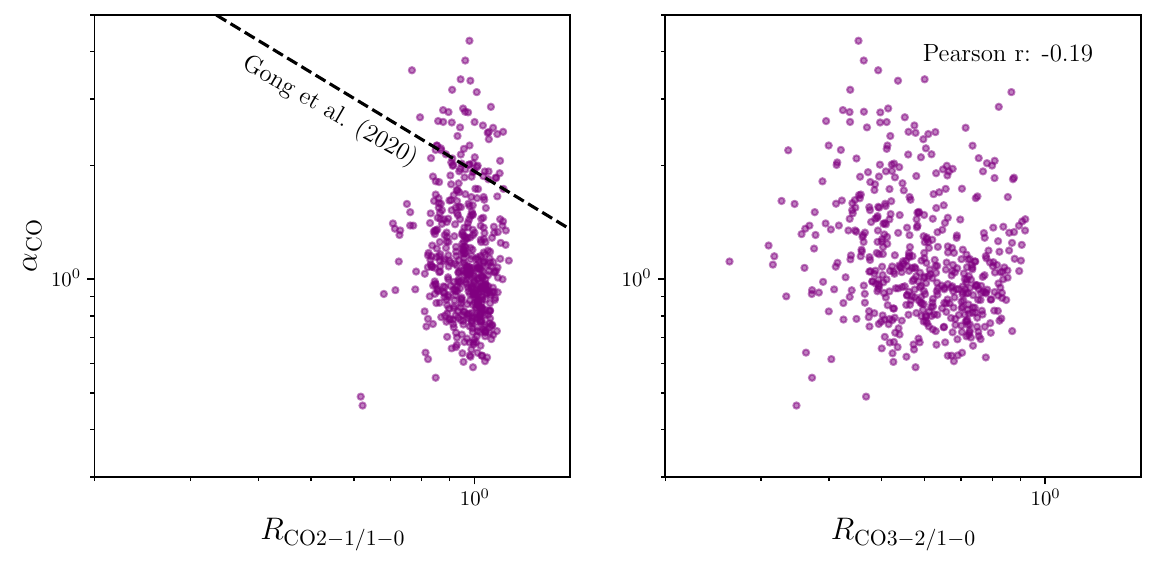}{0.9\textwidth}{}
	}
    \vspace{-2\baselineskip}
    \caption{$\alpha_{\mathrm{CO}}$ versus CO $J$=2-1/1-0 ratio $R_{\mathrm{CO 2{-}1/1{-}0}}$ (\textit{left}) and \cothree/1-0 ratio $R_{\mathrm{CO 3{-}2/1{-}0}}$ (\textit{right}) . The dashed line is the simulation fitting results from \citet{gong2020}. We can see there is no correlation between $\alpha_{\mathrm{CO}}$ and $R_{\mathrm{CO 2{-}1/1{-}0}}$ due to CO $J$=2-1 being thermalized, which saturates the ratio at values close to 1. We also do not see a significant anti-correlation between $\alpha_{\mathrm{CO}}$ and $R_{\mathrm{CO 3{-}2/1{-}0}}$, which suggests that CO line ratios are generally not a good tracer of $\alpha_{\mathrm{CO}}$ variation in starburst systems with large gas surface densities. }
    \label{fig:alphaCO_r21}
\end{figure*}

\citet{gong2020} proposed recently that the CO $J$=2-1/1-0 ratio $R_{\mathrm{CO 2{-}1/1{-}0}}$ can be used as a tracer of $\alpha_{\mathrm{CO}}$. For individual GMCs in virial equilibrium, we would expect the relation \citep{gong2020}
\begin{equation}
\alpha_{\mathrm{CO}} \propto 
\begin{cases} 
T_{\mathrm{ex}}(\text{CO 1-0})^{-1/2}, & \mathrm{low\ density} \\ 
\frac{\sqrt{n}}{T_{\mathrm{kin}}}, & \mathrm{high\ density},
\end{cases}
\end{equation}
In the low volume density regime, we have an anti-correlation between $\alpha_{\mathrm{CO}}$ and $T_{\mathrm{exc}}$. Since higher $T_{\mathrm{exc}}$ will directly lead to a higher $R_{\mathrm{CO 2{-}1/1{-}0}}$ ratio, we would expect an anti-correlation between $\alpha_{\mathrm{CO}}$ and  $R_{\mathrm{CO 2{-}1/1{-}0}}$. In Fig. \ref{fig:alphaCO_r21}, we see no correlation between these two quantities, which is probably due to our limited range of $R_{\mathrm{CO 2{-}1/1{-}0}}$ values (Fig. \ref{fig:co_ratio}). Furthermore, as discussed in Section \ref{subsec:ratio}, the $R_{\mathrm{CO 2{-}1/1{-}0}}$ ratio of GMCs in the Antennae is close to 1, which is due to both lines being thermally excited in regions that are warm and optically thick. In this case, we no longer expect an anti-correlation between $\alpha_{\mathrm{CO}}$ and $R_{\mathrm{CO 2{-}1/1{-}0}}$. 

Since $R_{\mathrm{CO 2{-}1/1{-}0}}$ saturates to $\sim$ 1, it is likely the CO excitation conditions are traced by ratios between higher-$J$ CO lines, such as \cothree, and CO $J$=1-0. However, although $R_{\mathrm{CO 3{-}2/1{-}0}}$ shows a significantly larger range of values than $R_{\mathrm{CO 2{-}1/1{-}0}}$, there is only a very weak anti-correlation between $\alpha_{\mathrm{CO}}$ and $R_{\mathrm{CO 3{-}2/1{-}0}}$. 
The weak correlation between $\alpha_{\mathrm{CO}}$ and CO line ratios suggest that excitation conditions does not play a major role in the $\alpha_{\mathrm{CO}}$ variation within starburst mergers. This is consistent with the observational studies in \citet{teng2023} for the centers of spiral galaxies, where they find temperature only has a minor ($\sim$ 20\%) contribution to $\alpha_{\mathrm{CO}}$ variations. Simulations, such as \citet{bournaud2015}, also find that $\alpha_{\mathrm{CO}}$ has weak or no correlation with high-$J$ CO to CO $J$=1-0 line ratios in starburst mergers.

We can also see from Fig. \ref{fig:alphaCO_r21} that our modeled $\alpha_{\mathrm{CO}}$ is mostly below the simulation prediction from \citet{gong2020}. This could be due to other environmental factors. For example, given a fixed $R_{\mathrm{CO 2{-}1/1{-}0}}$ ratio, the Antennae has higher GMC surface density (and higher CO $J$=1-0 intensity) than those in Milky Way like simulations, which could contribute to the lower $\alpha_{\mathrm{CO}}$. 

\subsection{Summary}

\begin{figure*}
\gridline{
    \leftfig{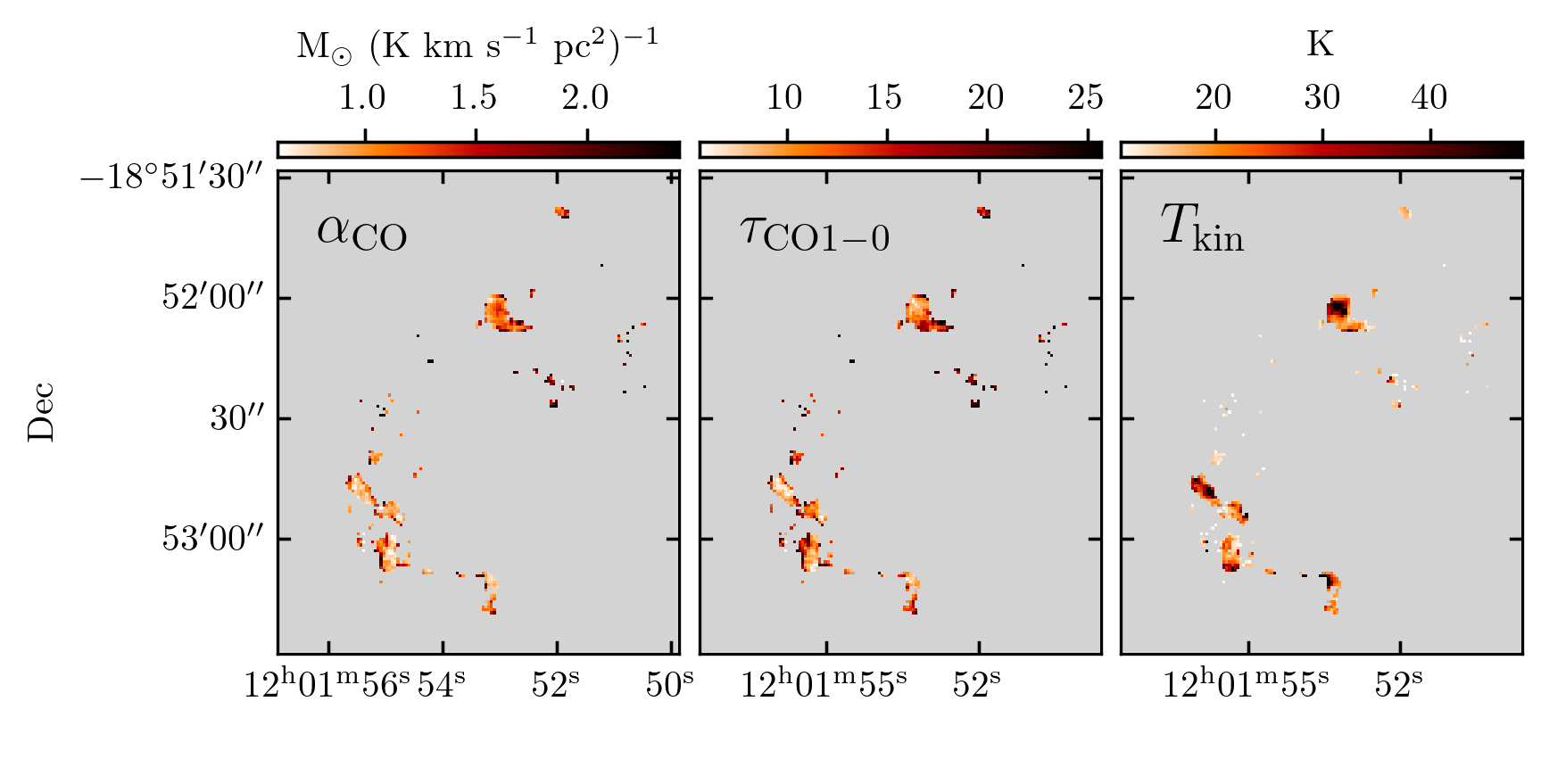}{0.9\textwidth}{}
	}
\vspace{-4\baselineskip}
\gridline{\leftfig{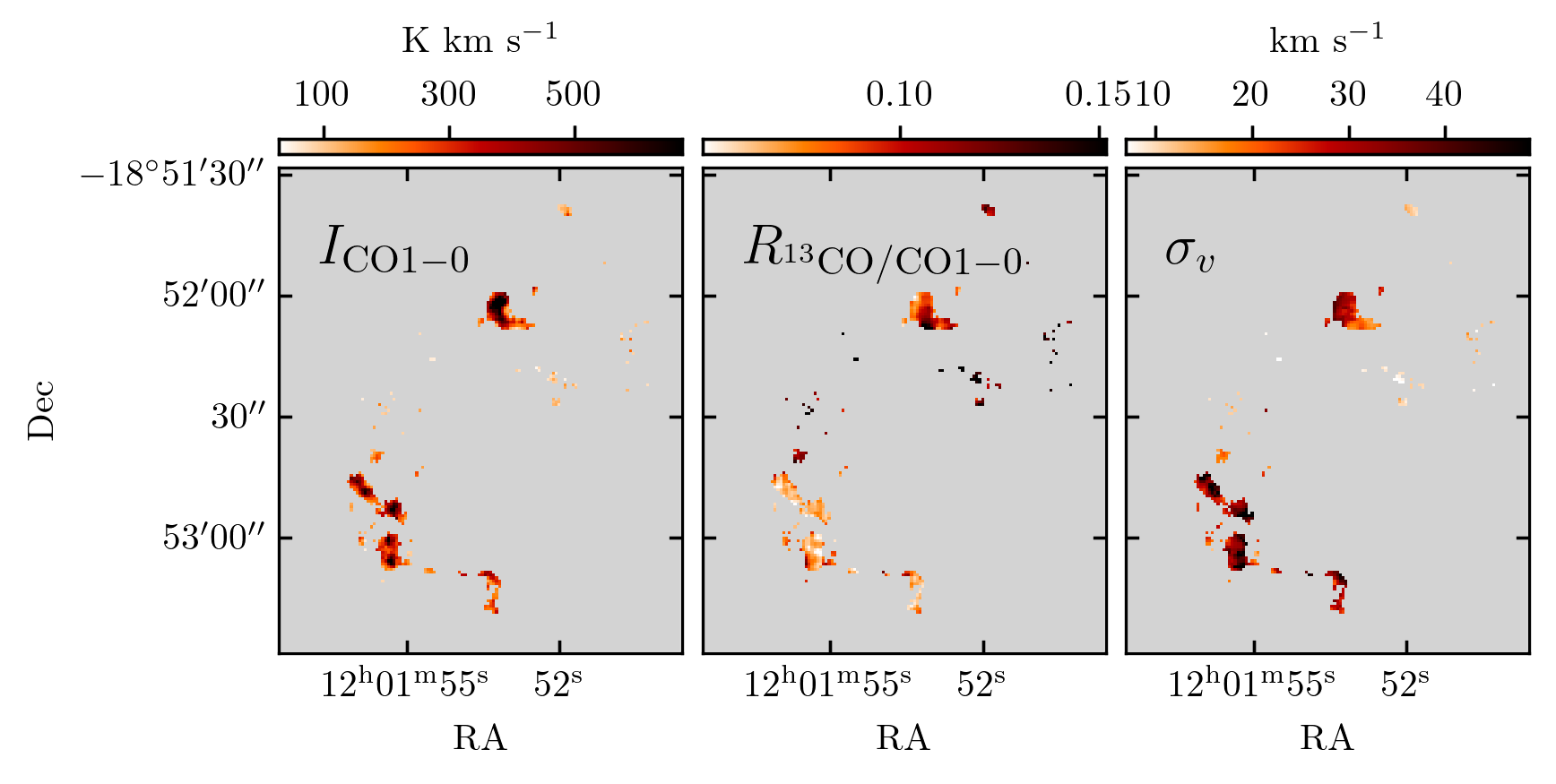}{0.9\textwidth}{}}
\vspace{-2\baselineskip}
    \caption{(\textit{Top}) The derived maps of $\alpha_{\mathrm{CO}}$, \taucoone and $T_{\mathrm{kin}}$ at 150 pc resolution. (\textit{Bottom}) The three observables that have strong correlations with $\alpha_{\mathrm{CO}}$. From left to right are CO $J$=1-0 integrated intensity ($I_{\mathrm{CO1-0}}$), $^{13}$CO/CO $J$=1-0 ratio ($R_{^{13}\mathrm{CO/CO}1-0}$) and CO $J$=1-0 velocity dispersion ($\sigma_v$).    
    }
    \label{fig:alphaCO_map}
\end{figure*}

The maps of $\alpha_{\mathrm{CO}}$, $\tau_{\mathrm{CO 1-0}}$ and $T_{\mathrm{kin}}$ are shown in Fig. \ref{fig:alphaCO_map}. As discussed in Section \ref{subsubsec:alphaCO_tau},  $\alpha_{\mathrm{CO}}$ is tightly correlated with $\tau_{\mathrm{CO 1-0}}$ so the two maps look quite similar. In contrast, $\alpha_{\mathrm{CO}}$ only has a weak anti-correlation with $T_{\mathrm{kin}}$ (Pearson coefficient of -0.17), which suggests that an increase in temperature is not the major factor in reducing $\alpha_{\mathrm{CO}}$ values in starburst mergers. This finding is consistent with studies of the centers of normal spiral galaxies \citep{teng2022, teng2023}. 

The bottom panels of Fig. \ref{fig:alphaCO_map} show the three observables that can be used to trace $\alpha_{\mathrm{CO}}$ variation. In theory, $R_{^{13}\mathrm{CO/CO}1-0}$ reflects the optical depth variation and hence the $\alpha_{\mathrm{CO}}$ variation (see Section \ref{subsubsec:alphaCO_tau}). The increase in $\sigma_v$ also acts in reducing the CO $J$=1-0 optical depth (Eq. \ref{eq:optical_depth}), and therefore reducing the $\alpha_{\mathrm{CO}}$ value. The strong anti-correlation between $\alpha_{\mathrm{CO}}$ and $I_{\mathrm{CO1-0}}$ is less well-explained from the theoretical side. It is likely that the CO-bright regions (with high $I_{\mathrm{CO1-0}}$) have higher temperature and lower optical depth  (due to large velocity dispersion), which act together to reduce the $\alpha_{\mathrm{CO}}$. We note that the $\alpha_{\mathrm{CO}}$ versus $I_{\mathrm{CO1-0}}$ relation has the shallowest slope among all the three relations, which suggests $\alpha_{\mathrm{CO}}$ is least sensitive to $I_{\mathrm{CO1-0}}$ variation. 

In this section, we have given several parameterized relations to infer $\alpha_{\mathrm{CO}}$ from direct observables. These relations, although capturing the same trends, are notably different from $\alpha_{\mathrm{CO}}$ prescriptions from previous simulations and observations. We note that most previous studies at cloud-scale resolution have been focused on molecular gas in normal spiral galaxies. Therefore, it is possible that the difference is caused by different galactic environments. We recommend our relation be used to infer $\alpha_{\mathrm{CO}}$ in starburst systems, such as U/LIRGs. To test our relation in a broader range of starburst environments, we need to expand our cloud-scale RADEX modeling to a larger sample of starburst U/LIRGs. 

\section{Constraining absolute $\alpha_{\mathrm{CO}}$ Values}

\subsection{Constraining the  $x_{\mathrm{CO}}$ abundance ratio with the dust mass}
\label{subsec:dust_xco}

From the dust continuum map, we calculate the dust mass as \citep{wilson_luminous_2008}
\begin{equation}
M_{\mathrm{dust}} = 74220 S_{880} D^2 \frac{\exp(17/T_{\mathrm{dust}})}{\kappa} 
\end{equation}
where $S_{880}$ is the flux in Jy, $D$ is the distance in Mpc, $T_{\mathrm{dust}}$ is the dust temperature in Kelvin and $\kappa$ is the dust opacity in g$^{-1}$ cm$^{2}$. We chose 0.9 g$^{-1}$ cm$^{2}$ as the fiducial value for $\kappa$ \citep{wilson_luminous_2008} but note that $\kappa$ can be a factor of 2 higher in starburst systems \citep{wilson2014}. The dust surface density for a given pixel can then be calculated as 
\begin{equation}
\begin{split}
\Sigma_{\mathrm{dust}} &= M_{\mathrm{dust}} / (1.1331 B_{\mathrm{FWHM}}^2) \\
                       &= 2.9 I_{880} D^2 \frac{\exp(17/T_{\mathrm{dust}})}{\kappa} \left(\frac{B_{\mathrm{FWHM}}}{150\ \mathrm{pc}}\right)^{-2}
\end{split}
\end{equation}
where $I_{880}$ is the intensity in Jy/beam and $B_{\mathrm{FWHM}}$ is the FWHM of the round beam in pc. We adopt the dust temperature from \citet{klaas2010} for pixels in each defined subregion (Fig. \ref{fig:dust_350GHz}). For subregion A1a which does not have a temperature measurement, we assume the temperature to be the same as the overall dust temperature derived from the integrated fluxes of the entire galaxy \citep{klaas2010}. 

\begin{figure}
\centering
\gridline{
    \fig{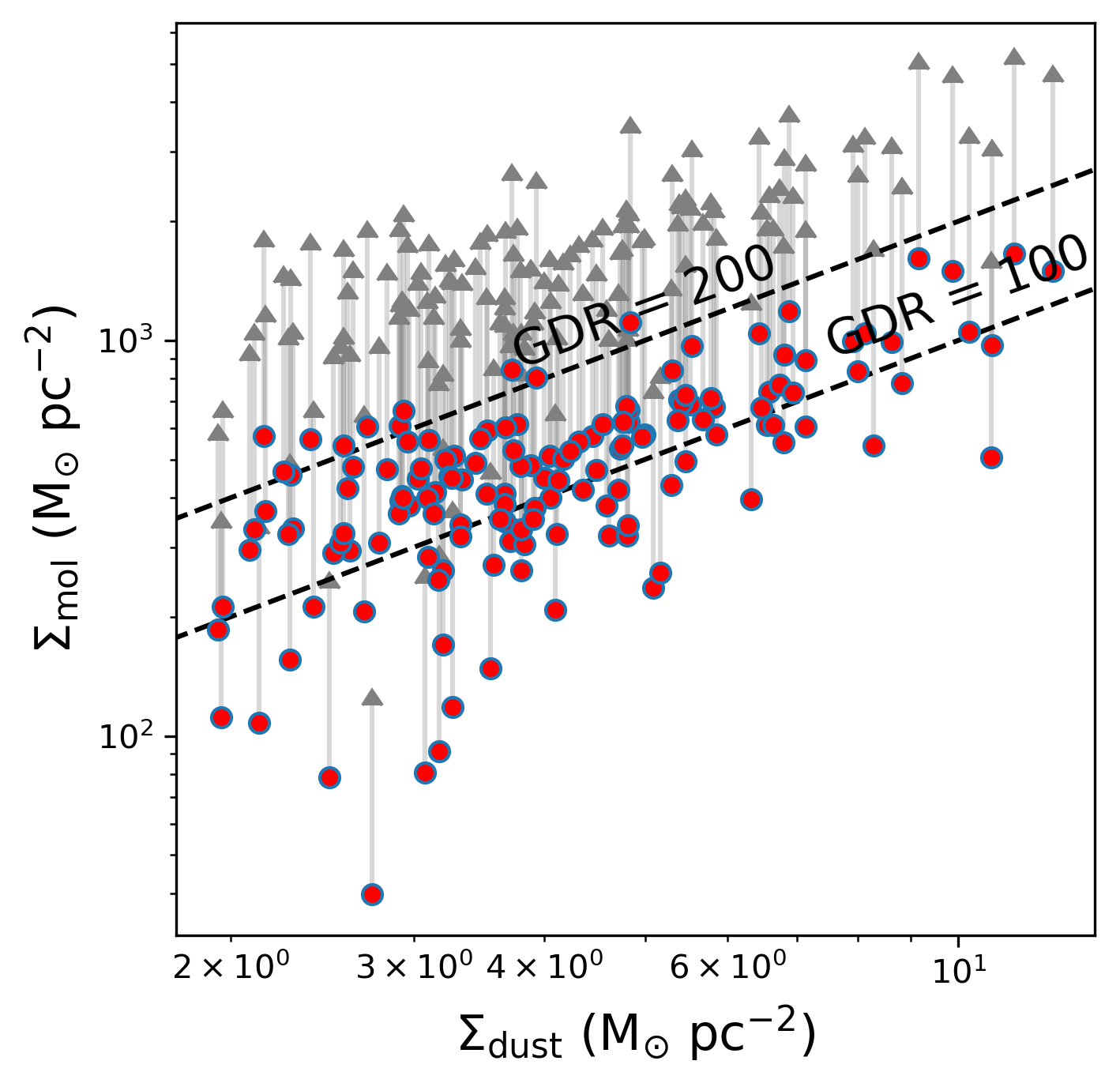}{0.4\textwidth}{}
	}
    \vspace{-2\baselineskip}
    \caption{Molecular gas versus dust surface density. \Sigmol is calculated using our modeled $\alpha_{\mathrm{CO}}$ values. Red circles and gray arrows indicate \Sigmol calculated assuming $x_{\mathrm{CO}}$ = 3e-4 and 1e-4 respectively. The dashed lines indicate constant gas-to-dust ratio (GDR). We can see that $x_{\mathrm{CO}}$ = $3 \times 10^{-4}$ gives us more realistic GDR values.
    }
    \label{fig:xco_GDR}
\end{figure}

The comparison of \Sigmol and $\Sigma_{\mathrm{dust}}$ is shown in the left panel of Fig. \ref{fig:xco_GDR}. We can see that a high $x_{\mathrm{CO}}$ value is favored in order to get a reasonable gas-to-dust ratio (GDR) below 200. Furthermore, since our modeling is targeting the densest GMCs, we would expect them to have GDR values closer to dense gas values of 50 -- 100 instead of more diffuse medium values of 200 \citep{remy2017}. If we instead assume a constant Milky Way $\alpha_{\mathrm{CO}}$ value, we would get GDR values of 300 -- 400 \citep[consistent with GDR values in][who adopted the Milky Way $\alpha_{\mathrm{CO}}$]{klaas2010}. 

An alternative explanation is that the Antennae might actually have a high GDR values. 
However, \citet{gunawardhana2020} derive the metallicity map for the Antennae and find most of regions in the Antennae have solar metallicity. Therefore, we would not expect the Antennae to have abnormally high GDR values.

The dust mass we calculate is affected by systematic uncertainties from parameters, such as dust temperature ($T_{\mathrm{dust}}$) and opacity ($\kappa$). The dust temperature calculated in \citet{klaas2010} is $\sim$ 20 K, while typical (U)LIRGs have temperatures of 20 -- 40 K \citep{dunne_dust_2022}. However, if we assume the higher dust temperature, we would obtain an even higher GDR value given a fixed $x_{\mathrm{CO}}$ value. The same logic also applies to dust opacity $\kappa$. \citet{wilson2014} suggest $\kappa$ can be two times higher in starburst systems than in the Milky Way. The higher $\kappa$ will lead to lower dust mass and hence higher GDR value given a fixed $x_{\mathrm{CO}}$. Therefore, increasing dust temperature and dust opacity does not help with reducing $x_{\mathrm{CO}}$ value, which further justifies our $x_{\mathrm{CO}}$ choice of 3$\times 10^{-4}$.

\subsection{GMC dynamical states in the Antennae}

\begin{figure}
\centering
\gridline{
    \fig{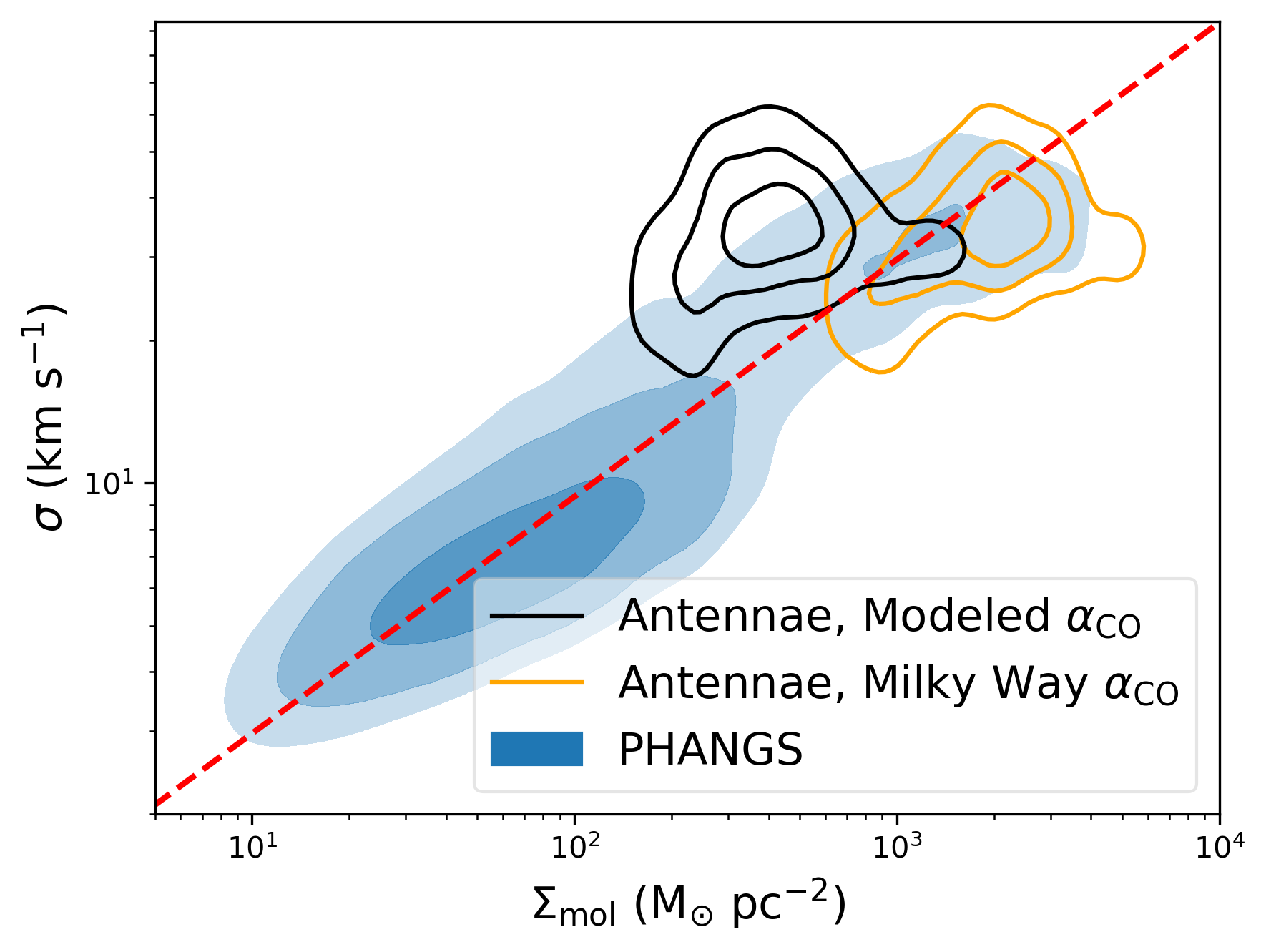}{0.45\textwidth}{}
	}
    \vspace{-2\baselineskip}
    \caption{Velocity dispersion versus gas surface density contours for PHANGS galaxies (blue shaded) and the Antennae using varying $\alpha_{\mathrm{CO}}$ from this work (black) and a constant Milky Way $\alpha_{\mathrm{CO}}$ value (orange). The red dashed line marks the position of the median value of \alphavir for PHANGS galaxies of 3.1 at 150 pc resolution \citep{sun_molecular_2020}. Our modeled $\alpha_{\mathrm{CO}}$ results suggest that GMCs in the Antennae are less gravitationally bound than GMCs in the PHANGS galaxies.   }
    \label{fig:vdep_Sigmol}
\end{figure}

As discussed in \citet{he2023}, the variation of $\alpha_{\mathrm{CO}}$ can lead to an uncertainty of a factor of 4 for GMC virial parameter (\alphavir) measurements, hence affecting our estimates of GMC dynamical states in galaxy mergers. With our modeled $\alpha_{\mathrm{CO}}$, we can put a more accurate constraint on GMC dynamical states in the Antennae. With $x_{\mathrm{CO}}$ = $3 \times 10^{-4}$, most of our modeled $\alpha_{\mathrm{CO}}$ are close to U/LIRG values of 1.1 $\mathrm{M_{\odot}\ (K\ km\ s^{-1}\ pc^{2})^{-1}}$, which is different from previous works that suggest that the Antennae should have a Milky Way $\alpha_{\mathrm{CO}}$ of 4.3 $\mathrm{M_{\odot}\ (K\ km\ s^{-1}\ pc^{2})^{-1}}$ \citep[e.g.][]{wilson_mass_2003, zhu2003, schirm_herschel-spire_2014}. This conclusion in general will increase the \alphavir of the Antennae, which suggests GMCs in the Antennae are less gravitationally bound than we might expect. As shown Fig. \ref{fig:vdep_Sigmol}, our modeled $\alpha_{\mathrm{CO}}$ results suggest GMCs in the Antennae are more turbulent, which is consistent with the simulation prediction from \citet{he2023}. 


Fig. \ref{fig:vdep_Sigmol} also shows that our modeled $\alpha_{\mathrm{CO}}$ gives a similar contour shape as adopting a constant Milky Way $\alpha_{\mathrm{CO}}$. This similarity suggests that the absolute value of the average $\alpha_{\mathrm{CO}}$ matters more than the relative $\alpha_{\mathrm{CO}}$ variation within the galaxy in determining the overall GMC dynamical states in the Antennae. 

\citet{wilson_mass_2003} first suggested a Milky Way $\alpha_{\mathrm{CO}}$ value based on comparison between virial mass and CO $J$=1-0 luminosity. However, GMCs in the Antennae are not necessarily in virial equilibrium. \citet{schirm_herschel-spire_2014} also suggest a typical $\alpha_{\mathrm{CO}}$ value of $\sim$ 7 $\mathrm{M_{\odot}\ (K\ km\ s^{-1}\ pc^{2})^{-1}}$ based on two-component LVG modeling assuming  $x_{\mathrm{CO}} = 3 \times 10^{-5}$. They choose this $x_{\mathrm{CO}}$ value so that the hot component of gas from CO LVG modeling is comparable to the mass derived from infrared H$_2$ emission. However, \citet{harrington_turbulent_2021} suggest that the high-$J$ CO transitions that are used to constrain the second component might come from cold dense gas instead of hot diffuse gas, and hence not necessarily trace the same gas component as the H$_2$ emission. If \citet{schirm_herschel-spire_2014} adopt our $x_{\mathrm{CO}}$ choice of $3 \times 10^{-4}$, they will get a similar $\alpha_{\mathrm{CO}}$ value ($\sim$ 0.7) as ours. 

\section{Modeled CO-to-H$_2$ conversion factor at kpc scales}

\subsection{Modeling setup}

In modeling $\alpha_{\mathrm{CO}}$ at kpc scales, we generally follow the same procedure as in Section \ref{subsec:RADEX_general}, except that we sample the beam filling factor in log space ($\log \Phi_{\mathrm{bf}}$ from -3 to 0 with step of 0.1). This is because we expect the beam filling factor at kpc scales could be much lower than 0.1. For the $\alpha_{\mathrm{CO}}$ modeling results, we also exclude pixels with one or more modeled quantities at the edge of our parameter space (as in Appendix \ref{app:alphaCO}), specifically, pixels with $X_{\mathrm{12/13, 1dmax}} \leq 30$, $\log T_{\mathrm{kin, 1dmax}} < 1.1$ or $\log T_{\mathrm{kin, 1dmax}} > 2$. 

\subsection{$\alpha_{\mathrm{CO}}$ comparison at GMC and kpc scales}
\label{subsec:GMC_kpc_comparison}

Most previous studies on constraining $\alpha_{\mathrm{CO}}$ have been done at kpc scales \citep[e.g.][]{papadopoulos_molecular_2012}. Previous simulations \citep[e.g][]{narayanan2012} argue that $\alpha_{\mathrm{CO}}$ should be a scale-free parameter down to cloud scale. In other words, the large-scale $\alpha_{\mathrm{CO}}$ should equal the luminosity weighted cloud-scale $\alpha_{\mathrm{CO}}$. However, our observed line intensity maps at kpc-scale resolution are the convolution product of line intensity maps at cloud-scale resolution and the convolving beam set by the resolution of the telescope, 
and \alphaco modeling based on the convoluted data is not a linear process as averaging \alphaco weighted by the \coone luminosities, and hence could lead to different \alphaco results. Furthermore, RADEX modeling generally assumes a uniform spherical structure, which is more applicable to individual GMCs than an ensemble of GMCs at kpc-scale resolution. Therefore, it is important to compare $\alpha_{\mathrm{CO}}$ modeling results at both scales to quantify the impact of these factors. 

\begin{figure*}
\gridline{
    \fig{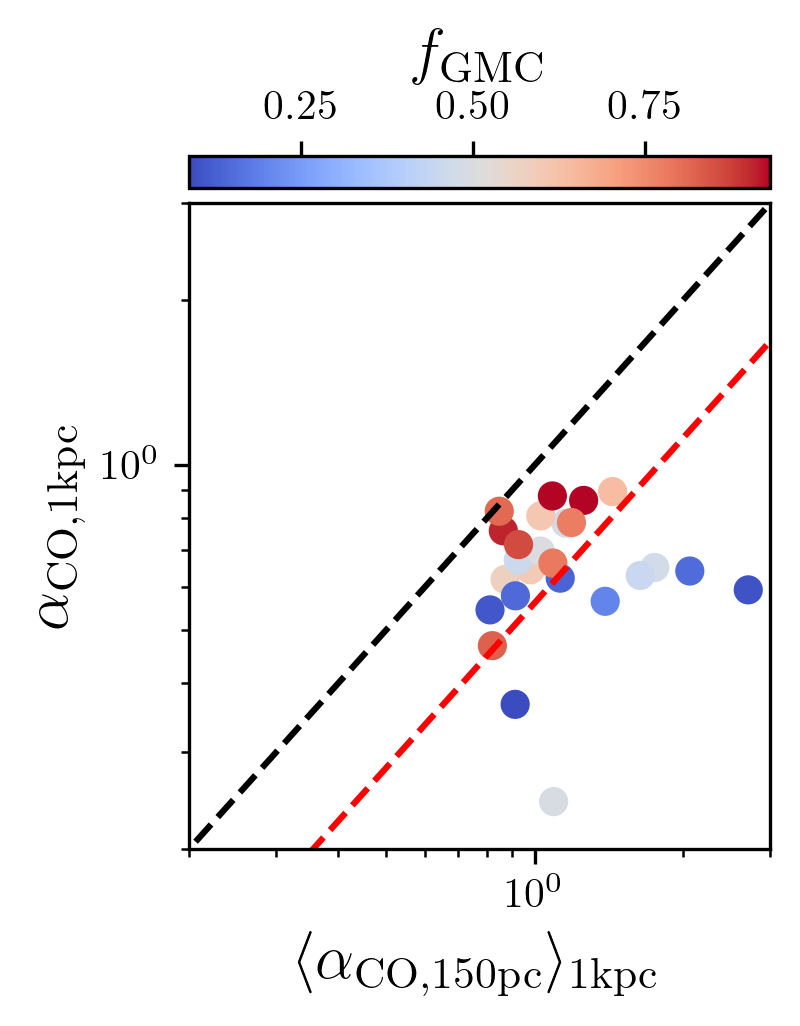}{0.33\textwidth}{}
    \fig{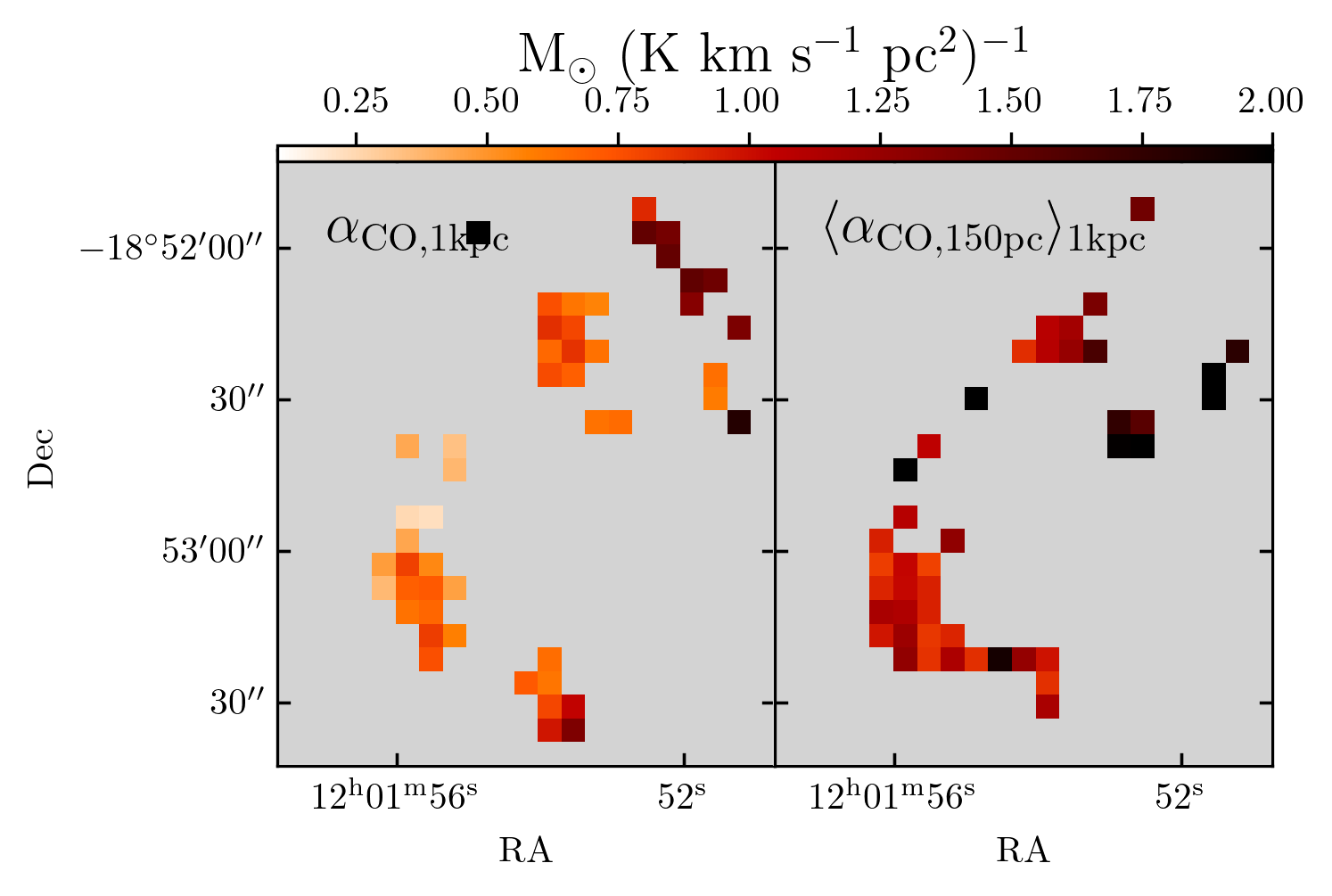}{0.66\textwidth}{}
}
    \vspace{-2\baselineskip}
    \caption{(\textit{Left}) The comparison between $\alpha_{\mathrm{CO}}$ at kpc scale ($\alpha_{\mathrm{CO, 1kpc}}$) and  $\alpha_{\mathrm{CO}}$ at 150 pc scale intensity averaged over kpc scale pixels ($\langle\alpha_{\mathrm{CO, 150pc}}\rangle_{\mathrm{1kpc}}$). Pixels are color coded by the fraction of CO $J$=1-0 emission at kpc scale that comes from GMC scales for each pixel ($f_{\mathrm{GMC}}$, see text for detailed description). The black dashed line is the one-to-one line while the red dashed line is the proportional fit to the data (coefficient of 0.56). We can see that the higher fraction of the GMC-scale emission at kpc scale roughly corresponds to higher $\alpha_{\mathrm{CO, 1kpc}}$/$\langle\alpha_{\mathrm{CO, 150pc}}\rangle_{\mathrm{1kpc}}$ ratio, which suggests a diffuse gas component might play a role in bringing down the $\alpha_{\mathrm{CO, 1kpc}}$. 
    (\textit{Right}) Maps of $\alpha_{\mathrm{CO, 1kpc}}$ and $\langle\alpha_{\mathrm{CO, 150pc}}\rangle_{\mathrm{1kpc}}$. }
    \label{fig:alphaCO_kpc}
\end{figure*}

We generate the $\alpha_{\mathrm{CO}}$ map at kpc scale by applying the same procedure as in Section \ref{sec:RADEX_modeling} using the combo masked CO moment maps at 1 kpc resolution generated from the PHANGS-ALMA pipeline. To match the $\alpha_{\mathrm{CO}}$ maps at 150 pc and 1 kpc resolution, we regrid the 150 pc map to have the same pixel size as the 1 kpc map. For each regridded 150 pc map pixel, $\alpha_{\mathrm{CO}}$ is calculated as the CO $J$=1-0 intensity averaged value of the smaller pixels that are associated with the regridded pixel, 
\begin{equation}
\langle\alpha_{\mathrm{CO, 150pc}}\rangle_{\mathrm{1kpc}} = \frac{\sum \alpha_{\mathrm{CO, 150pc}} L^{\mathrm{150 pc}}_{\mathrm{CO(1-0)}}}{\sum  L^{\mathrm{150 pc}}_{\mathrm{CO(1-0)}}}, 
\end{equation}
where $L^{\mathrm{150 pc}}_{\mathrm{CO(1-0)}}$ is the CO $J$=1-0 luminosity of each pixel for 150 pc resolution map. We note that this calculation only includes pixels with robust $\alpha_{\mathrm{CO}}$ results (i.e. with detection of all five lines). The blank pixels are excluded and GMCs with highest surface densities have the highest weight. The comparison between $\alpha_{\mathrm{CO}}$ at both scales is shown in Fig. \ref{fig:alphaCO_kpc}. 

We can see that $\alpha_{\mathrm{CO}}$ values at kpc scale are about 60\% of the CO $J$=1-0 intensity averaged $\alpha_{\mathrm{CO}}$ values at 150 pc scale. It is possible that kpc-scale CO emission includes a diffuse component that is not detected at GMC scales. This diffuse component would be warmer and more luminous and hence would have lower $\alpha_{\mathrm{CO}}$ values \citep[][]{schirm_herschel-spire_2014, kamenetzky_warm_2017}. To quantify this effect, we calculate the fraction of the emission at kpc scale that comes from dense GMCs. For each kpc-scale pixel, the total flux of GMC emission is calculated by summing up fluxes of all pixels in the 150 pc resolution map associated with the kpc-scale pixel and with valid $\alpha_{\mathrm{CO}}$ values. The GMC fraction is
\begin{equation}
f_{\mathrm{GMC}} = \frac{\sum L^{\mathrm{150 pc}}_{\mathrm{CO(1-0)}}}{L^{\mathrm{1 kpc}}_{\mathrm{CO(1-0)}}}, 
\end{equation}
where $L^{\mathrm{1 kpc}}_{\mathrm{CO(1-0)}}$ is the CO $J$=1-0 luminosity for pixels in the 1 kpc resolution map. 
We note this sum also only includes pixels with $^{13}$CO detections, and hence is more representative of GMCs with high volume/surface density. GMCs excluded from the summing generally have low surface densities (or equivalently low \coone luminosities). Therefore, we expect them to have higher \alphaco values based on the \alphaco versus $I_{\mathrm{CO 1-0}}$ trend (Fig. \ref{fig:alphaCO_Ico}). In other words, our cloud-scale averaged \alphaco values are biased towards lower values. In the left panel of Fig. \ref{fig:alphaCO_kpc}, we color code the data points with $f_{\mathrm{GMC}}$. Data points with high fraction ($\sim$ 100\%) are closer to the 1-to-1 line, which is consistent with our expectation. 
We also calculate the Pearson correlation coefficient between $\alpha_{\mathrm{CO, 1kpc}}/\langle\alpha_{\mathrm{CO, 150pc}}\rangle_{\mathrm{1kpc}}$ and $f_{\mathrm{GMC}}$ and find a strong correlation between these two quantities with the coefficient of 0.62. This correlation suggests that an additional component of diffuse gas at kpc scale brings down the overall $\alpha_{\mathrm{CO}}$ value. 


However, it is still under debate whether the diffuse component of molecular gas has higher or lower $\alpha_{\mathrm{CO}}$ compared to cold dense gas in GMCs. Liszt, Pety \& Lucas (2010) show that $\alpha_{\mathrm{CO}}$ is relatively constant among different molecular gas components in our Milky Way. They suggest this constant $\alpha_{\mathrm{CO}}$ should be attributed to the offsetting effects of lower CO abundances with respect to H$_2$ ($x_{\mathrm{CO}}$) and a large $I_{\mathrm{CO}}$/\Nco ratio in low extinction gas. Recent studies by \citet{ramambason2023} further suggest that the abundance factor plays a more dominant role and hence actually increases $\alpha_{\mathrm{CO}}$ in diffuse molecular gas. We note that our modeling does not have the ability to constrain $x_{\mathrm{CO}}$. Therefore, the $\alpha_{\mathrm{CO}}$ we modeled is proportional to the  \Nco/$I_{\mathrm{CO}}$ ratio and hence does not reflect any $\alpha_{\mathrm{CO}}$ change due to CO abundance variation. For example, if the diffuse gas actually has lower $x_{\mathrm{CO}}$, we would expect the actual $\alpha_{\mathrm{CO}}$ at kpc scale to be higher than our modeled values.

\subsection{$\alpha_{\mathrm{CO}}$ dependence at kpc scale}

\begin{figure*}
\centering
\gridline{
    \fig{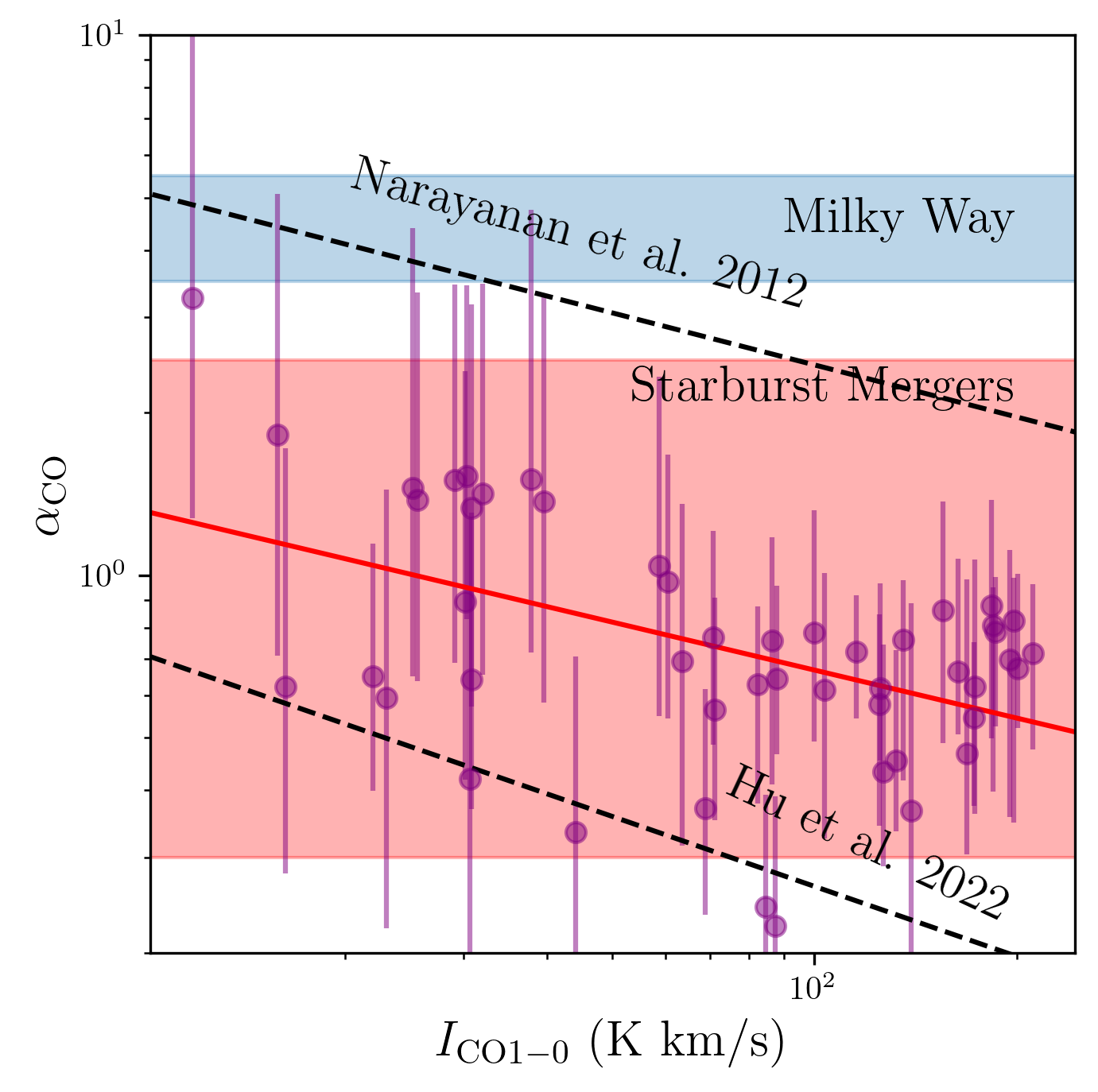}{0.45\textwidth}{}
    \hspace{-\baselineskip}
    \fig{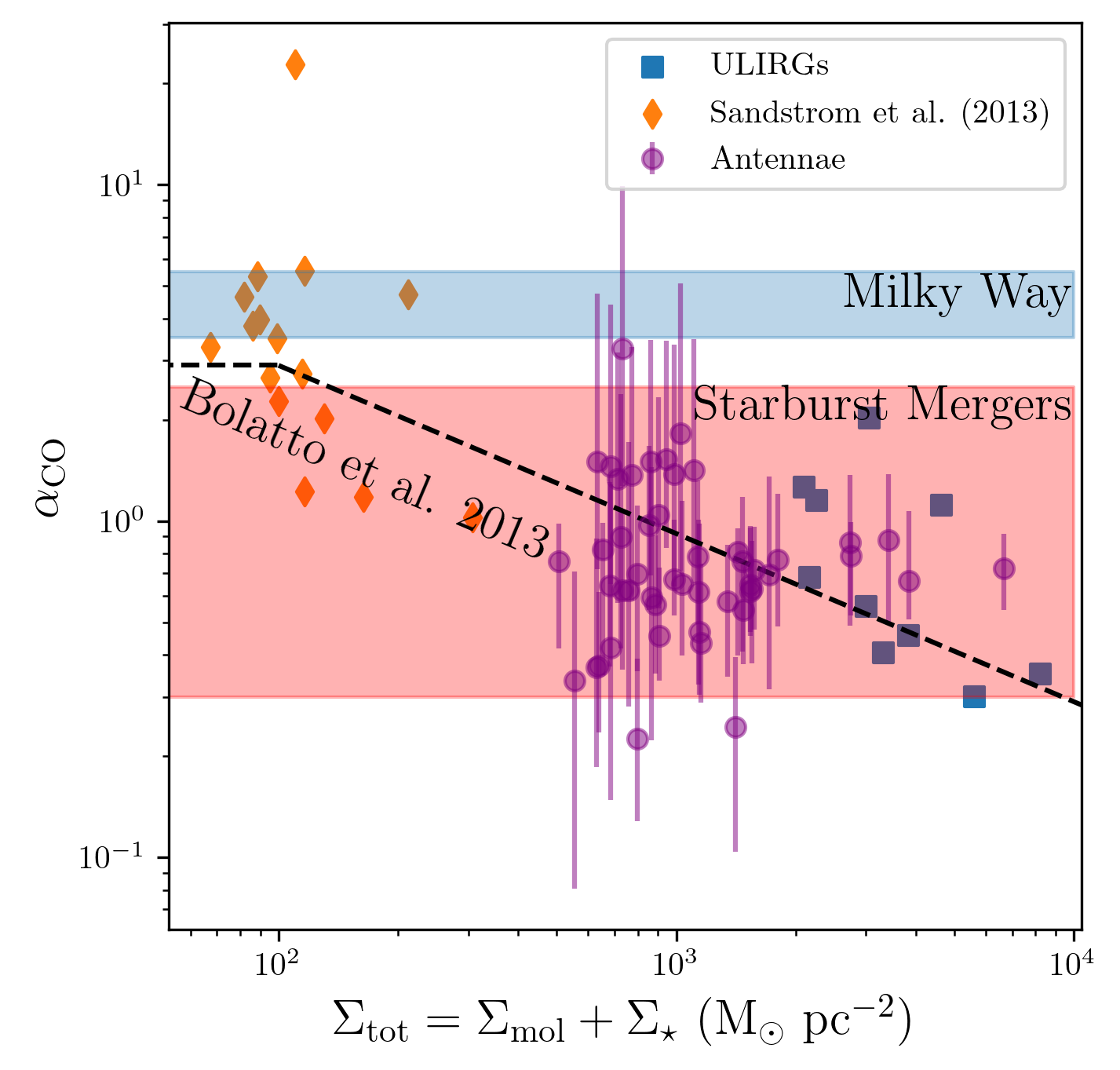}{0.45\textwidth}{}
	}
    \vspace{-2\baselineskip}
    \caption{(\textit{Left}) Modeled $\alpha_{\mathrm{CO}}$ versus CO $J$=1-0 integrated intensity $I_{\mathrm{CO(1-0)}}$ at 1 kpc resolution. The error bar marks the 16th and 84th values in the modeled $\alpha_{\mathrm{CO}}$ 1D distribution.
    The red solid line is the power law fit to the data while the dashed lines are the simulation predicted relations from \citet{narayanan2012} and \citet{hu2022}, respectively. We can see our fit relation has similar slope as the simulation predictions. (\textit{Right}) Modeled $\alpha_{\mathrm{CO}}$ versus total surface density for the Antennae (purple), normal spiral galaxies (orange diamonds) and ULIRGs (blue squares). The dashed line is the prescription from \citet{bolatto_co--h_2013}. We can see that the absolute value of $\alpha_{\mathrm{CO}}$ in the Antennae is generally consistent with the prediction of this prescription.}
    \label{fig:alphaCO_kpc_dependence}
\end{figure*}

Due to limited resolution and sensitivity, previous studies have been mostly focused on kpc-scale $\alpha_{\mathrm{CO}}$ calibrations. Besides metallicity, the two most widely used kpc-scale observables to calibrate $\alpha_{\mathrm{CO}}$ are CO $J$=1-0 intensity \citep{narayanan2012} and total galactic disk surface density \citep[stellar plus gas,][]{bolatto_co--h_2013}. However, while these prescriptions generally capture galaxy-to-galaxy variations, they are less well tested for $\alpha_{\mathrm{CO}}$ variations within individual galaxies, specifically for distant starburst galaxies that are hard to resolve. We test these two dependencies in the Antennae using our kpc-scale $\alpha_{\mathrm{CO}}$ data (Fig. \ref{fig:alphaCO_kpc_dependence}). 

The left panel of Fig. \ref{fig:alphaCO_kpc_dependence} shows $\alpha_{\mathrm{CO}}$ versus $I_{\mathrm{CO 1-0}}$ in comparison with simulation predictions. We can see a negative correlation between these two quantities with slope (-0.3) close to the two simulation predictions \citep[-0.32 for ][-0.43 for \citealt{hu2022}]{narayanan2012}. We also see offsets between observed and simulation predicted absolute values. We note that the simulation by \citet{hu2022} is focused on a typical kpc-size disk region in the Milky Way with a maximum $I_{\mathrm{CO 1-0}}$ of 1 K km s$^{-1}$ at kpc scales. Our observed $I_{\mathrm{CO 1-0}}$ is clearly out of this range. We also note that our $\alpha_{\mathrm{CO}}$ versus $I_{\mathrm{CO 1-0}}$ correlation at GMC scale is in relatively good agreement with the \citet{hu2022} prediction (Fig. \ref{fig:alphaCO_Ico}). Therefore, the discrepancy in the kpc-scale $\alpha_{\mathrm{CO}}$ versus $I_{\mathrm{CO 1-0}}$ correlation might be caused by differences in GMC beam filling factor at kpc scales. If the number of GMCs inside the simulated kpc box were increased, we would expect higher kpc-scale $I_{\mathrm{CO 1-0}}$ for a given $\alpha_{\mathrm{CO}}$, which would bring the simulation predicted relation rightward to become more aligned with our observed $\alpha_{\mathrm{CO}}$. On the other hand, the simulation from \citet{narayanan2012} gives larger $\alpha_{\mathrm{CO}}$ than our modeled results. We note that $\alpha_{\mathrm{CO}}$ at kpc scale in \citet{narayanan2012} is calculated as the CO $J$=1-0 intensity averaged value of $\alpha_{\mathrm{CO}}$ of individual clouds, which is similar to what we did to calculate $\langle\alpha_{\mathrm{CO, 150pc}}\rangle_{\mathrm{1kpc}}$ in Section \ref{subsec:GMC_kpc_comparison}. Therefore, the discrepancy might be due to our inclusion of a diffuse molecular gas component that brings down the $\alpha_{\mathrm{CO}}$ values. It is also possible that the true $x_{\mathrm{CO}}$ is slightly lower than $3 \times 10^{-4}$ and the actual $\alpha_{\mathrm{CO}}$ in the Antennae might be higher than our derived values. 

The right panel of Fig. \ref{fig:alphaCO_kpc_dependence} shows $\alpha_{\mathrm{CO}}$ versus the total surface density compared with the empirical relation by \citet{bolatto_co--h_2013}. The absolute $\alpha_{\mathrm{CO}}$ values are generally consistent with the theoretical expectation but with a large scatter. We also do not see a significant correlation between these two quantities for the Antennae alone (Pearson coefficient of -0.11) or the Antennae and other U/LIRGs combined (Pearson coefficient of -0.16). However, the large uncertainty (factor of $\sim$3) in our modeled $\alpha_{\mathrm{CO}}$ values might act to smear out the trend. 

\section{Conclusions} \label{sec:discussion}

In this paper, we have constrained the spatial variation of the CO-to-H$_2$ conversion, $\alpha_{\mathrm{CO}}$, in the Antennae merger at both GMC and kpc scales based on high-resolution ALMA CO and $^{13}$CO lines. Our main conclusions are summarized below.

\begin{itemize}
    \item The CO $J$=2-1/1-0 ($\sim$ 1) and \cothree/1-0 ($\sim$ 0.7) ratios in the Antennae are significantly higher than the commonly observed ratio in normal spiral galaxies \citep{leroy2021, wilson2012, leroy2022}. These large ratios suggest that molecular gas in this starburst system has higher volume density and/or kinetic temperature compared to normal spiral galaxies. \item The $^{13}$CO/CO $J$=1-0 and $^{13}$CO/CO $J$=2-1 ratios ($\sim$ 0.1) in the Antennae are similar to those in normal spiral galaxies \citep{cormier_full-disc_2018} but larger than typical U/LIRGs \citep[$\sim 0.02$][]{brown2019}. The difference in the isotopologue ratios between the Antennae and U/LIRGs is likely due to optical depth effects as U/LIRGs might be undergoing stronger stellar feedback from more intense starburst activity, which disperses the molecular gas and reduces its optical depth.  
    \item We have derived the first resolved $\alpha_{\mathrm{CO}}$ map for the Antennae down to GMC scale ($\sim$ 150 pc). We find that $\alpha_{\mathrm{CO}}$ has a significant anti-correlation with GMC integrated intensity, $I_{\mathrm{CO 1-0}}$, which is consistent with simulation predictions \citep{gong2020,hu2022}. This supports the argument that $\alpha_{\mathrm{CO}}$ has a continuous dependence on $I_{\mathrm{CO 1-0}}$ instead of a bimodal distribution among normal spiral and starburst galaxies. 
    \item We find that $\alpha_{\mathrm{CO}}$ has a strong, tight linear correlation with the CO optical depth. This suggests that $\alpha_{\mathrm{CO}}$ variations in starburst systems are mainly driven by optical depth variations rather than kinetic temperature variations, which is consistent with $\alpha_{\mathrm{CO}}$ studies in normal galaxy centers \citep{teng2022, teng2023}\textbf{}. We also find a relatively tight correlation between $\alpha_{\mathrm{CO}}$ and the $^{13}$CO/CO $J$=1-0 ratio. This correlation is consistent with our expectation that the $^{13}$CO/CO $J$=1-0 ratio can probe the molecular gas optical depth. The scatter in the $\alpha_{\mathrm{CO}}$ versus $^{13}$CO/CO $J$=1-0 ratio is mainly driven by the varying [$\mathrm{CO}$]/[$\mathrm{^{13}CO}$] abundance ratio. 
    \item We find that $\alpha_{\mathrm{CO}}$ is also tightly related to the GMC dynamical state. The strong anti-correlation between $\alpha_{\mathrm{CO}}$ and $\sigma_v$ has a slope consistent with theoretical prediction of -0.5. This result is consistent with previous LVG studies on U/LIRGs \citep[e.g.][]{papadopoulos_molecular_2012}, which suggested that the low $\alpha_{\mathrm{CO}}$ values in these systems are mainly caused by GMCs with large velocity dispersion and hence being out of virial equilibrium.    
    \item We compare our modeled gas surface density with 345 GHz dust continuum. Our comparison shows that our chosen $[\mathrm{CO}]$/$[\mathrm{H_2}]$ abundance ratio  $x_{\mathrm{CO}} = 3 \times 10^{-4}$ gives us reasonable gas-to-dust ratios of $\sim$ 100. Given this abundance ratio choice, we would expect most $\alpha_{\mathrm{CO}}$ values in the Antennae are close to the typical U/LIRG value of 1.1 $\mathrm{M_{\odot}\ (K\ km\ s^{-1}\ pc^{2})^{-1}}$. This $\alpha_{\mathrm{CO}}$ will put most GMCs in the Antennae out of virial equilibrium, which is consistent with simulation predictions of GMCs in starburst mergers \citep{he2023}.   
    \item We compare luminosity weighted GMC-scale $\alpha_{\mathrm{CO}}$ values averaged at kpc resolution with $\alpha_{\mathrm{CO}}$ values directly derived from kpc-resolution data. Our comparison shows that kpc-scale $\alpha_{\mathrm{CO}}$ from LVG modeling is about 60\% of the averaged value of $\alpha_{\mathrm{CO}}$ at 150 pc scale. We think that the lower $\alpha_{\mathrm{CO}}$ at kpc scale might be due to a diffuse warm component that has intrinsically lower $\alpha_{\mathrm{CO}}$. 
    \item We also explore the dependence of the modeled $\alpha_{\mathrm{CO}}$ at kpc scales on various observables. We find that the kpc-scale $\alpha_{\mathrm{CO}}$ shows a similar anti-correlation with CO intensity as predicted by simulations \citep[e.g.][]{narayanan2012, hu2022}. We also tested the anti-correlation between $\alpha_{\mathrm{CO}}$ and total surface density, $\Sigma_{\mathrm{tot}}$, as suggested in \citet{bolatto_co--h_2013}. We find $\alpha_{\mathrm{CO}}$ of the Antennae lies along the trend with normal spiral galaxies and U/LIRGs with the absolute values consistent with the prescription prediction. We do not find a significant anti-correlation between $\alpha_{\mathrm{CO}}$ and $\Sigma_{\mathrm{tot}}$ for the Antennae alone, which could be due to large uncertainties in our modeled $\alpha_{\mathrm{CO}}$ values.  
\end{itemize}


We thank the referee for thoughtful comments and constructive suggestions. This paper makes the uses of the following ALMA data: \\
ADS/JAO.ALMA \# 2018.1.00272.S \\
ADS/JAO.ALMA \# 2021.1.00439.S. 
ALMA is a partnership of ESO (representing its member states), NSF (USA), and NINS (Japan), together with NRC (Canada), MOST and ASIAA (Taiwan), and KASI
(Republic of Korea), in cooperation with the Republic of Chile. The Joint ALMA Observatory is operated by ESO, AUI/NRAO, and NAOJ. The National Radio Astronomy Observatory is a facility of the National Science Foundation operated
under cooperative agreement by Associated Universities, Inc. This work is based in part on observations made with the Spitzer Space Telescope, which was operated by the Jet Propulsion Laboratory, California Institute of Technology under a contract with NASA. The research of CDW and ER
is supported by grants from the Natural Sciences and Engineering Research Council of Canada (NSERC), and also for CDW the Canada Research Chairs program. JS acknowledges support by the National Aeronautics and Space Administration (NASA) through the NASA Hubble Fellowship grant HST-HF2-51544 awarded by the Space Telescope Science Institute (STScI), which is operated by the Association of Universities for Research in Astronomy, Inc., under contract NAS~5-26555.  

\section*{Data Availability}

The code to generate RADEX modeling grids and perform Bayesian analysis on multiple CO and its isotopologue lines is available on both Zenodo (\url{https://zenodo.org/records/10845936}) and GitHub$^{\ref{github}}$. The modeled maps of gas physical properties and CO-to-H$_2$ conversion factor are available from the corresponding author on request. 

%

\vspace{5mm}
\facilities{ALMA, Spitzer}


\software{astropy \citep{the_astropy_collaboration_astropy_2013, Astropy_2018, Astropy_2022}, CASA \citep{CASA_2022}
          }


\bibliography{references}{}
\bibliographystyle{aasjournal}

\setcounter{figure}{0} 
\renewcommand{\thefigure}{A\arabic{figure}} 

\begin{figure*}
\centering
\gridline{
    \fig{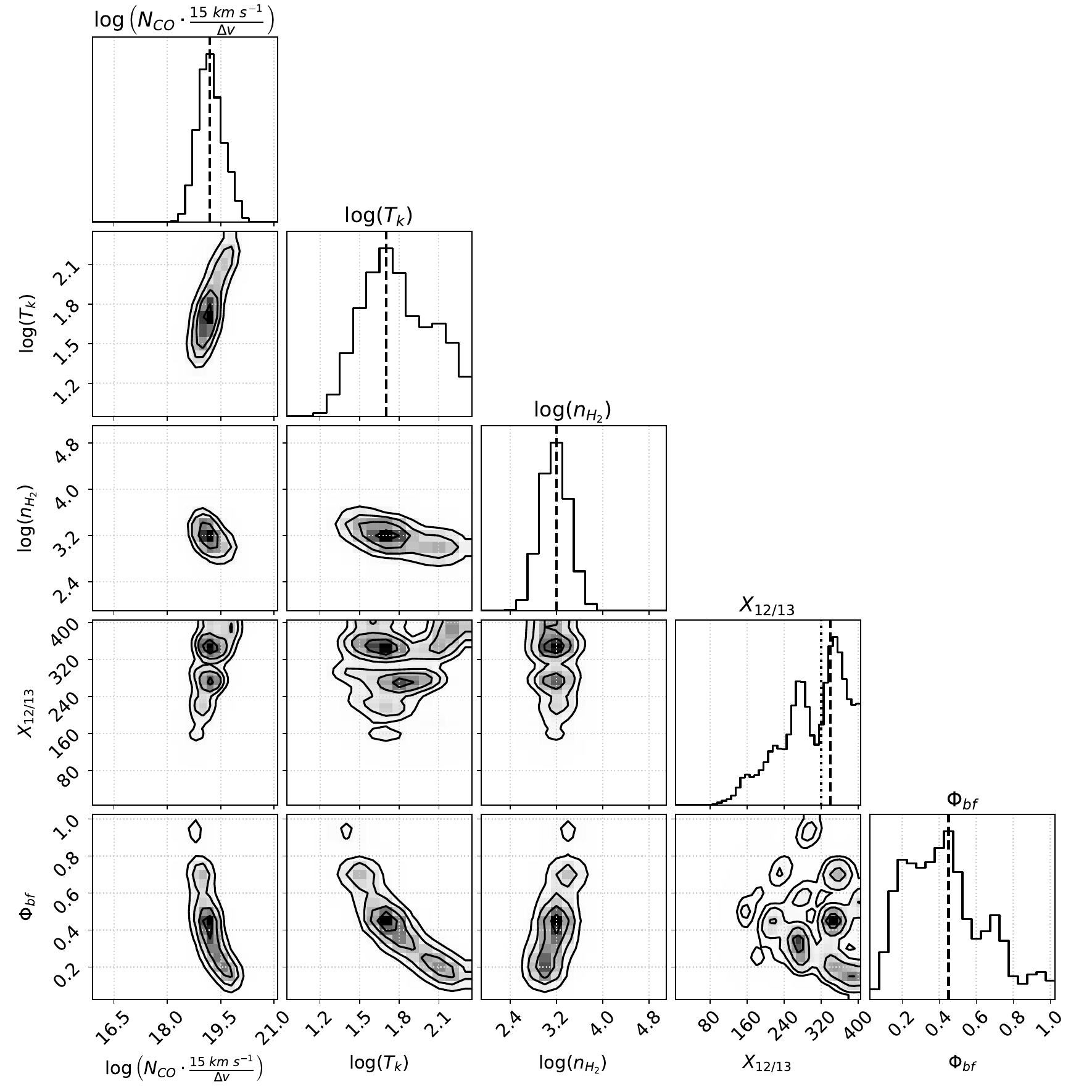}{0.9\textwidth}{}
	}
    \vspace{-2\baselineskip}
    \caption{Corner plot of modeled RADEX physical properties for the north nucleus pixel (Fig. \ref{fig:co10_moments}). From the left to right (also top to bottom) are the scaled CO column density (cm$^{-2}$), kinetic temperature (K), hydrogen volume density (cm$^{-3}$), $[\mathrm{CO}]/[^{13}\mathrm{CO}]$ abundance ratio and beam filling factor. The dashed and dotted lines mark the maximal and median of the 1D distribution for each quantity. }
    \label{fig:RADEX_marginalize}
\end{figure*}


\restartappendixnumbering
\appendix

\section{Bayesian Analyses in Our RADEX Modeling}
\label{app:RADEX_stats}

\begin{table}
\centering
\caption{Solutions for the Modeled Parameter}
\label{tab:solution}
\begin{threeparttable}
\begin{tabularx}{0.35\textwidth}{cc}
    \hline \hline
     Solution & Condition \\
     \hline
     Bestfit &  $P(\vec{\theta}_{\mathrm{Bestfit}}|\vec{I}_{\mathrm{obs}}) = \max P(\vec{\theta}|\vec{I}_{\mathrm{obs}})$  \\
     1DMax & $P(\theta_{i, \mathrm{1DMax}}|\vec{I}_{\mathrm{obs}}) = \max P(\theta_i|\vec{I}_{\mathrm{obs}})$   \\
     Neg1Sig & $\int^{\theta_{i, \mathrm{Neg1Sig}}} d\theta_i P(\theta_i|\vec{I}_{\mathrm{obs}}) = 0.16$ \\
     Median & $\int^{\theta_{i, \mathrm{Median}}} d\theta_i P(\theta_i|\vec{I}_{\mathrm{obs}}) = 0.5$ \\
     Pos1Sig & $\int^{\theta_{i, \mathrm{Pos1Sig}}} d\theta_i P(\theta_i|\vec{I}_{\mathrm{obs}}) = 0.84$ \\ 
     \hline
\end{tabularx}
\end{threeparttable}
\end{table}

We follow the Bayesian likelihood analyses in \citet{teng2022} to characterize the probability density function (PDF) for the five varying parameters. For each pixel, the code calculates the $\chi^2$ by comparing the modeled line intensities with the measured line intensities. Note that in \citet{teng2022}, they compare modeled and measured integrated line intensities. While they assumed a fixed velocity FWHM that is representative of their observed region, this may introduce the issue that the fixed velocity FWHM (15 \velu) in the modeling is inconsistent with varying velocity FWHMs measured for different pixels in the observations \citep[see Appendix B in][for futher discussion]{teng2023}. Since our modeled intensity corresponds to linewidth of 15 \velu, we need to rescale modeled intensity by multiplying the ratio of measured linewidth to the fixed linewidth of 15 \velu for each pixel.  Then we calculate the $\chi^2$ matrix as
\begin{equation}
\chi^2 (\vec{\theta}) = \sum_{i=1}^{N=5} \frac{I^{\mathrm{mod, scaled}}_i(\vec{\theta})-I^{\mathrm{obs}}_i}{\sigma_i^2}
\end{equation}
where $\vec{\theta}$ represents each modeled parameter set of $(n_{\mathrm{H_2}}, T_{\mathrm{kin}}, N_{\mathrm{CO}}/\Delta v, X_{12/13}, \Phi_{\mathrm{bf}})$, $I_i^{\mathrm{mod, scaled}} = I_i^{\mathrm{mod}} \frac{\Delta v}{15 \mathrm{km s^{-1}}}$ represents the scaled modeled integrated intensity for each line and  $I_i^{\mathrm{obs}}$ represents the integrated intensities from observations, $\sigma_i$ the measurement uncertainty for $I_i^{\mathrm{obs}}$ of each line and $N$ specifies the number of lines used for the modeling. For each pixel value, we calculate the posterior probability distribution function across the 5D model parameter space as
\begin{equation}
P(\vec{\theta}|\vec{I}_{\mathrm{obs}}) = \frac{1}{Q} \exp(-\chi^2/2)
\end{equation}
where $Q^2 = (2\pi)^5 \prod_{i} \sigma_i^2$ is the normalization coefficient. From the 5D distribution, we can calculate the 'Bestfit' set of modeled parameters with maximal $P(\vec{\theta}|\vec{I}_{\mathrm{obs}})$. We can also calculate the marginalized 1D probability distribution for each individual modeled parameter by integrating the 5D $P(\vec{\theta}|\vec{I}_{\mathrm{obs}})$ over the rest of parameter space. The equation for the 1D marginalized distribution is 
\begin{equation}
P(\theta_i|\vec{I}_{\mathrm{obs}}) = \idotsint \limits_{j \neq i} d\theta_j \ P(\vec{\theta}|\vec{I}_{\mathrm{obs}}) 
\end{equation}
where $\theta_i$ is the one modeled parameter that we want to calculate the 1D marginalized distribution and $\theta_j$ are the rest of modeled parameters.

We show the 1D and 2D probability distribution for each modeled quantity in Fig. \ref{fig:RADEX_marginalize} for one of the pixels in north nucleus (marked as red dot in Fig. \ref{fig:co10_moments}). For most quantities, the median of the 1D probability distribution corresponds well with the `1DMax' value of the distribution, which suggests that our modeling recovers the most probable solution. One exception is $X_{12/13}$, which has double peaks in its 1D distribution. This is a common behavior for most of our pixels, which suggests that the $X_{12/13}$ value is less well constrained. Therefore, for all of the modeled physical quantities except for $X_{12/13}$, we adopt the median value as our modeled solution. For $X_{12/13}$, we instead use the `1DMax' value as the modeled solution. 

\section{Modeling of the CO-to-H2 conversion factor}
\label{app:alphaCO}

\setcounter{figure}{0} 
\renewcommand{\thefigure}{B\arabic{figure}} 

\begin{figure}
\centering
\gridline{
    \fig{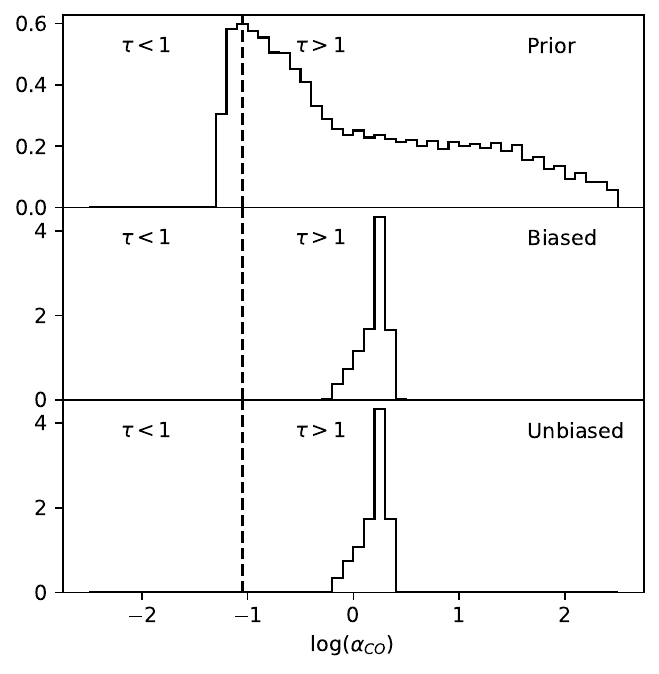}{0.4\textwidth}{}
	}
    \vspace{-2\baselineskip}
    \caption{Marginalized $\alpha_{\mathrm{CO}}$ distribution for the north nucleus pixel (Fig. \ref{fig:co10_moments}). (\textit{Top}) The prior distribution of $\alpha_{\mathrm{CO}}$ calculated by attributing uniform weighting to every set of parameters. (\textit{Middle}) The histogram of $\alpha_{\mathrm{CO}}$ posterior distribution without sampling bias correction. (\textit{Bottom}) The histogram of $\alpha_{\mathrm{CO}}$ posterior distribution after correcting for the sampling bias. To the right of the dashed line is where the $\alpha_{\mathrm{CO}}$ is not well sampled in our parameter space. This is also the line that roughly divides the optically thin and optically thick regime (see Eq. \ref{eq:alphaCO_tau}).}
    \label{fig:alphaCO_marginalize}
\end{figure}

We calculate the $\alpha_{\mathrm{CO}}$ 1D distribution by summing up all the probabilities of parameters in 5D space that yield a given $\alpha_{\mathrm{CO}}$ value, which is 
\begin{equation}
\begin{split}
P_{\mathrm{biased}}(\alpha_{\mathrm{CO}} | \vec{I}_{\mathrm{obs}}) &= P_{\mathrm{biased}}(\frac{N_{\mathrm{CO}}\Phi_{\mathrm{bf}}} {x_{\mathrm{co}}I^{\mathrm{mod, scaled}}_{\mathrm{CO(1-0)}}}  | \vec{I}_{\mathrm{obs}}) \\
&= \int\limits_{f(\Vec{\theta}) = \alpha_{\mathrm{CO}}} d\Vec{\theta} \ P(\vec{\theta}|\vec{I}_{\mathrm{obs}})
\end{split}
\end{equation}
where $f(\Vec{\theta})\equiv \frac{N_{\mathrm{CO}}\Phi_{\mathrm{bf}}} {x_{\mathrm{co}}I^{\mathrm{mod, scaled}}_{\mathrm{CO(1-0)}}}$. However, we need to note that $\alpha_{\mathrm{CO}}$ is not uniformly sampled in our 5D parameter space. To get the unbiased PDF, we calculate the normalized ratio of our biased $\alpha_{\mathrm{CO}}$ probability to the $\alpha_{\mathrm{CO}}$ prior probability in our sampling space, which is 
\begin{equation}
P(\alpha_{\mathrm{CO}} | \vec{I}_{\mathrm{obs}}) = \left[\frac{P_{\mathrm{biased}}(\alpha_{\mathrm{CO}} | \vec{I}_{\mathrm{obs}})}{P_{\mathrm{prior}}(\alpha_{\mathrm{CO}})} \right]_{\mathrm{norm}} 
\end{equation}
An example of the 3 $\alpha_{\mathrm{CO}}$ PDFs for one pixel is shown in Fig. \ref{fig:alphaCO_marginalize}. 

We find that some pixels have extremely small $\alpha_{\mathrm{CO}}$ values. As shown in Fig. \ref{fig:alphaCO_marginalize}, $\alpha_{\mathrm{CO}}$ values below 0.1 are not well sampled and the prior distribution has a huge drop below this threshold. Therefore, we exclude the pixels with median $\alpha_{\mathrm{CO}}$ value below 0.1. We also exclude $\alpha_{\mathrm{CO}}$ with large uncertainties ($\alpha_{\mathrm{CO, pos1sig}}/\alpha_{\mathrm{CO, neg1sig}} < 16$, which means uncertainty within a factor of 4). We further exclude pixels with 1dMax values of modeled quantities at the edge of our parameter space. Specifically, we apply the selection criterion to mask out pixels with $X_{\mathrm{12/13, 1dMax}} \leq 30$, $\log n_{\mathrm{H_2, 1dMax}} \geq 4.9$, $\log (N_{\mathrm{CO, 1dMax}} \times \frac{15 \mathrm{km s^{-1}}}{\Delta v}) \leq 16.1$ and $\log (T_{\mathrm{kin, 1dMax}}) < 1.1 $. We also exclude pixels with $\chi^2$ value greater than 10. 

\setcounter{figure}{0} 
\renewcommand{\thefigure}{C\arabic{figure}} 

\begin{figure*}
\vspace{-1\baselineskip}
\centering
\gridline{
    \fig{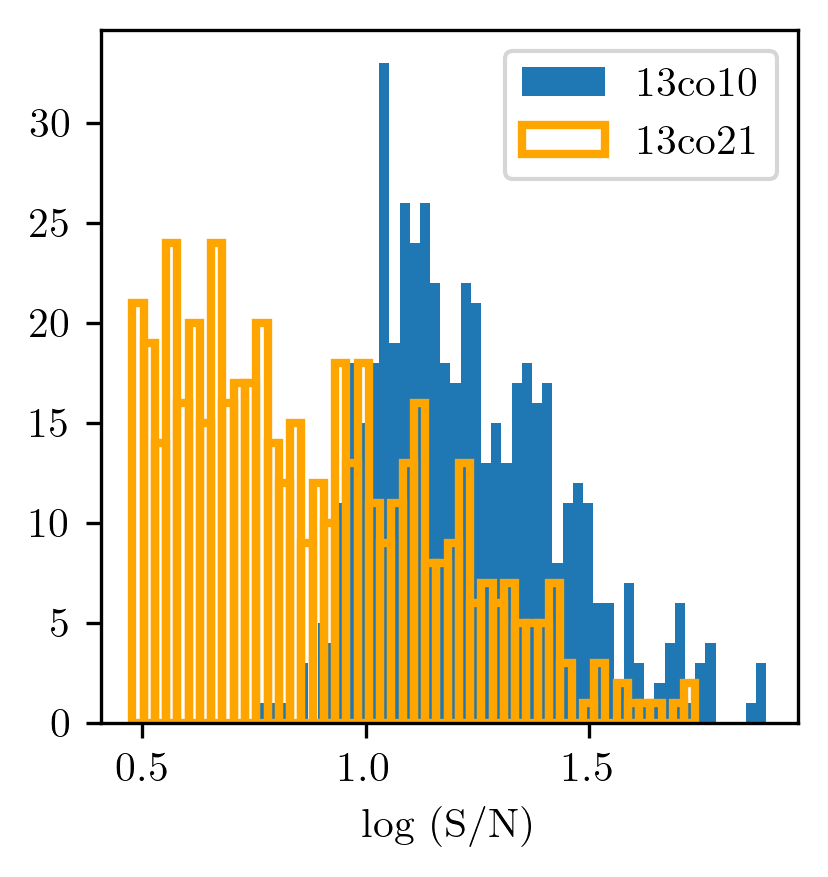}{0.3\textwidth}{}
    \fig{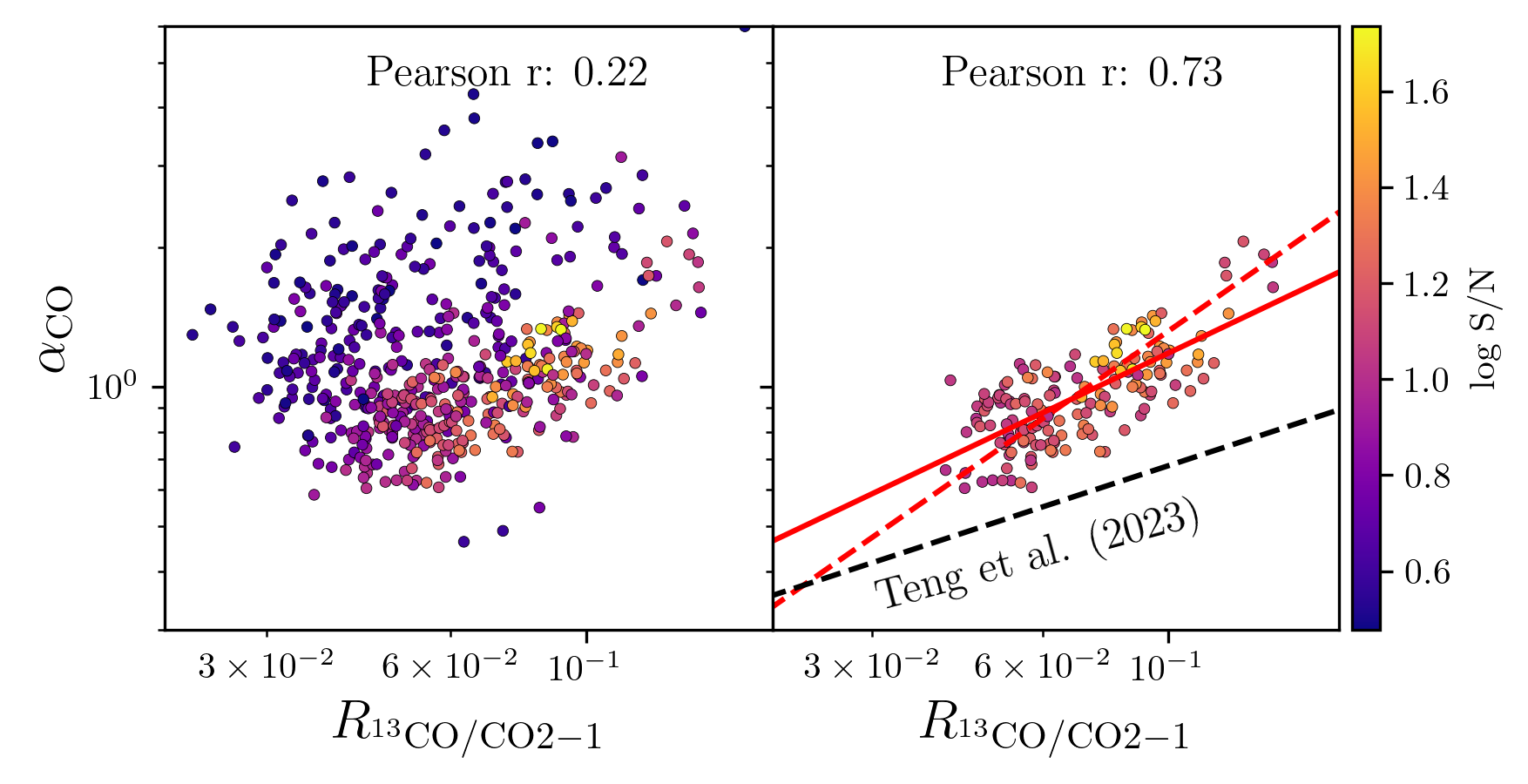}{0.62\textwidth}{}
	}
    \vspace{-2\baselineskip}
    \caption{(\textit{Left}) The logarithm of the S/N distribution of $^{13}$CO $J$=1-0 and $J$=2-1 for pixels with good $\alpha_{\mathrm{CO}}$ constraints. We can see most pixels have $^{13}$CO $J$=1-0 S/N greater than 10. In contrast, majority of pixels have $^{13}$CO $J$=2-1 S/N less than 10. (\textit{Right}) The $\alpha_{\mathrm{CO}}$ versus the $^{13}$CO/CO $J$=2-1 ratio for all pixels with good $\alpha_{\mathrm{CO}}$ constraints (left panel) and for pixels with $^{13}$CO $J$=2-1 S/N level greater than 10 (right panel). The red solid line is the power-law fit for the $\alpha_{\mathrm{CO}}$ versus $R_{\mathrm{^{13}CO/CO 2{-}1}}$ relation while the red dashed line is the fit for the $\alpha_{\mathrm{CO}}$ versus $R_{\mathrm{^{13}CO/CO 1{-}0}}$ relation. The black dashed line is the relation from \citet{teng2023}. We can see that by removing low S/N pixels, $\alpha_{\mathrm{CO}}$ also has a tight correlation with the $^{13}$CO/CO 2-1 ratio.}
    \label{fig:alphaCO_R21}
\end{figure*}

\section{$\alpha_{\mathrm{CO}}$ versus $^{13}$CO/CO $J$=2-1 ratio}
\label{sec:ratio_sensitivity}

In Fig. \ref{fig:alphaCO_tau}, we find a strong correlation between $\alpha_{\mathrm{CO}}$ and the $^{13}$CO/CO $J$=1-0 ratio, which is consistent with the literature findings in \citet{teng2022,teng2023}. However, we see that our $\alpha_{\mathrm{CO}}$ versus $R_{\mathrm{^{13}CO/CO 1{-}0}}$ relation has a steeper slope than the $\alpha_{\mathrm{CO}}$ versus $R_{\mathrm{^{13}CO/CO 2{-}1}}$ relation found in \citet{teng2022,teng2023}. One possible reason for the difference might be due to our different choice of excitation lines. To test this, we also plot the $\alpha_{\mathrm{CO}}$ versus $R_{\mathrm{^{13}CO/CO 2{-}1}}$ for the Antennae in Fig. \ref{fig:alphaCO_R21}. We can see that if we include all the data points from Fig. \ref{fig:alphaCO_tau}, we have at best a weak correlation between $\alpha_{\mathrm{CO}}$ and $R_{\mathrm{^{13}CO/CO 2{-}1}}$. We note that our $^{13}$CO $J$=2-1 observation has a much lower sensitivity than the $^{13}$CO $J$=1-0 observation, which results in a much larger scatter in the observed $R_{\mathrm{^{13}CO/CO 2{-}1}}$. In the left panel of Fig. \ref{fig:alphaCO_R21}, we show the S/N distribution for the $^{13}$CO $J$=1-0 and $J$=2-1 data. We can clearly see that most ($\sim$ 90\%) of the $^{13}$CO $J$=1-0 data has S/N level greater than 10 while less than half ($\sim$ 30\%) of the $^{13}$CO $J$=2-1 data achieve the same S/N level. In our $\alpha_{\mathrm{CO}}$ versus $R_{\mathrm{^{13}CO/CO 2{-}1}}$ plot color coded by the S/N level, we see most points that cause the scatter have S/N less than 10. If we remove those pixels with S/N smaller than 10, we see a much stronger correlation between $\alpha_{\mathrm{CO}}$ and $R_{\mathrm{^{13}CO/CO 2{-}1}}$ with a Pearson correlation coefficient of 0.7, similar to the coefficient of $\alpha_{\mathrm{CO}}$ versus $R_{\mathrm{^{13}CO/CO 1{-}0}}$. This suggests that we need $^{13}$CO data with high sensitivity in order to use the $^{13}$CO/CO ratio to calibrate $\alpha_{\mathrm{CO}}$. 

We perform the same power-law fit to the $\alpha_{\mathrm{CO}}$ versus $R_{\mathrm{^{13}CO/CO 2{-}1}}$ relation for pixels with S/N$>10$, as shown in the rightmost panel in Fig.  \ref{fig:alphaCO_R21}. The function is 
\begin{equation}
 \log \alpha_{\mathrm{CO}} = 0.58 (\pm 0.04) \log R_{\mathrm{^{13}CO/CO 2{-}1}} + 0.65 (\pm 0.05)     
\end{equation}
We also overlay the $\alpha_{\mathrm{CO}}$ versus $R_{\mathrm{^{13}CO/CO 1{-}0}}$ relation in the same figure panel. We can see that the $^{13}$CO/CO $J$=2-1 ratio gives a shallower slope than the $^{13}$CO/CO $J$=1-0 ratio. It is possible that the shallower slope is due to a larger fraction of $^{13}$CO $J$=2-1 lines being subthermally excited. In the subthermal case, the $^{13}$CO/CO ratio might also track the temperature/volume density variation as the $T_{\mathrm{ex}}$ for the two lines cannot be canceled out (see Eq. \ref{eq:R10_tau13}). Hence, the $^{13}$CO/CO $J$=2-1 ratio might have a less steep dependence on the optical depth, which is tightly correlated with $\alpha_{\mathrm{CO}}$. This could result in a shallower slope for the $\alpha_{\mathrm{CO}}$ versus $^{13}$CO/CO $J$=2-1 relation.

\end{document}